\documentclass[10pt,twocolumn,oneside,final]{IEEEtran}

\usepackage{cite}
\usepackage{graphicx}
\usepackage{amsmath}
\usepackage{times}
\usepackage{latexsym}
\usepackage{bm}
\usepackage{amssymb}
\usepackage[center]{caption2}
\usepackage{array}
\usepackage{setspace}
\usepackage{fancyhdr}
\usepackage{cite,graphicx,amsmath,amssymb}
\usepackage{citesort}
\usepackage{color}
\usepackage{multirow}
\usepackage{times}
\usepackage{amsmath}
\usepackage{amssymb}
\usepackage{amsfonts}
\usepackage{graphics}
\usepackage{amssymb}
\usepackage{booktabs}

\ifCLASSINFOpdf
\else
\fi

\makeatother

\begin{document}
\title{MSE-based Precoding for MIMO Downlinks in Heterogeneous Networks}
\author{Yongyu~Dai,~\IEEEmembership{Student Member,~IEEE,}
        Xiaodai~Dong,~\IEEEmembership{Senior~Member,~IEEE,}~and~Wu-Sheng~Lu,~\IEEEmembership{Fellow,~IEEE}
\thanks{
Y.~Dai, X.~Dong, and W.-S.~Lu are with the Department of Electrical and Computer Engineering, University of Victoria, Victoria, BC, Canada (email: yongyu@uvic.ca,~xdong@ece.uvic.ca,~wslu@ece.uvic.ca).
}}
\maketitle

%

\begin{abstract}
    Considering a heterogeneous network (HetNet) system consisting of a macro tier overlaid with a second tier of small cells (SCs), this paper studies the mean square error (MSE) based precoding design to be employed by the macro base station and the SC nodes for multiple-input multiple-output (MIMO) downlinks. First, a new sum-MSE of all users based minimization problem is proposed aiming to design a set of macro cell (MC) and SC transmit precoding matrices or vectors. To solve it, two different algorithms are presented. One is via a relaxed-constraints based alternating optimization (RAO) realized by efficient alternating optimization and relaxing non-convex constraints to convex ones. The other is via an unconstrained alternating optimization with normalization (UAON) implemented by introducing the constraints into the iterations with the normalization operation. Second, a separate MSE minimization based precoding is proposed by considering the signal and interference terms corresponding to the macro tier and the SCs separately. Simulation results show that the sum-MSE based RAO algorithm provides the best MSE performance among the proposed schemes under a number of system configurations. When the number of antennas at the macro-BS is sufficiently large, the MSE of the separate MSE-based precoding is found to approach that of RAO and surpass that of UAON. Together, this paper provides a suite of three new precoding techniques that is expected to meet the need in a broad range of HetNet environments with adequate balance between performance and complexity.
\end{abstract}

\IEEEpeerreviewmaketitle

\section{Introduction}\label{sec:introduction}

It is widely acknowledged that further improvements in network capacity are only possible by increasing the node deployment density~\cite{3gpp36.814,hetnet2011ltea}. On the other hand, deploying more macro tiers in already dense networks may be prohibitively expensive and result in significantly reduced cell splitting gains due to severe inter-cell interference~\cite{hetnet20113gpp}. Heterogeneous networks (HetNets) that embed a large number of low-power nodes into an existing macro network with the aim of offloading traffic from the macro cell to small cells has emerged as a viable and cost-effective way to increase network capacity~\cite{3gpp36.814,hetnet2011ltea,hetnet20113gpp,5G}.

In a typical HetNet consisting of a macro cell (MC) and several small cells (SCs), the MC serves its user equipments (UEs) in a large region by a high-power base station (BS), while each SC serves its UEs in its own coverage region by a low-power SC node if there is no cooperative transmission between the BSs and SCs\footnote{In this paper, SC is also utilized to denote the SC node for simplicity.}. Due to the large number of potential interfering nodes in the network, properly mitigating both the inter-cell and intra-cell multiuser interference is a crucial issue facing HetNet. Interference control (IC) for the interference networks recently has been intensively studied and applied in HetNet~\cite{Gesbert2010,Hong2013,CompHetNet2014,Zhu2011,Dai2013}, and the coordinated multi-point (CoMP) transmission is demonstrated to be an effective approach in \cite{Gesbert2010}, including joint processing (JP) and coordinated beamforming (CB). When the backhaul among the coordinated tiers is able to share both user data and channel state information (CSI), the CoMP-JP transmission is shown to provide high spectral efficiency~\cite{Hong2013,CompHetNet2014}. However, JP also introduces limitations for practical implementation due to its needs for high signaling overhead. On the other hand, with the BSs and SCs cooperated in the beamformer or precoder level, CB strategies only require the share of CSI in order to mitigate the cross-tier interference between the macro cell and co-channel deployed SCs. Reference \cite{Zhu2011} has implemented the cross-tier IC with CB based on a prioritized user selection scheme. Later, a joint selection based IC is presented to achieve more balanced performances between the macro cell UEs and the SC UEs~\cite{Dai2013}. Nevertheless, these schemes with closed-form expressions are only available in certain cases, such as a two-user multiple-input multiple-output (MIMO) interference channel.

In practical systems, the design of specific interference control schemes is subject to various criteria and constraints. Typically, interference control is formulated as problems that optimize certain system utility functions, which are directly associated with the UE rates or mean square error (MSE). Since the signal-to-noise ratio (SNR) is not so high in the practical wireless systems, especially at the cell edge, imperative performance improvement in the low and intermediate SNR region becomes a motivation in the IC scheme design. In \cite{Hui2010}, new MSE-based transceiver schemes are designed through efficient iterative algorithms for the peer to peer MIMO interference channel. In addition, source and relay precoding designs based on the MSE criterion in MIMO two-way relay systems are investigated in \cite{Wang20121,Wang20122}. Unfortunately, due to their differences in network architecture, they may not be employed directly into the HetNet systems, where there are hierarchical nodes including BS and SCs and each of them can transmit to multiple users.

To the best knowledge of the authors, there are no MSE-based precoding schemes for HetNet in the literature. In this paper, we develop three new MSE-based precoding schemes for MIMO downlinks in HetNet systems consisting of a macro tier overlaid with a second tier of SCs. Collectively, the proposed precoding schemes form a design toolbox that is expected to cover a wide spectrum of system needs ranging from superior precoding performance for systems with sufficient computing power to non-iterative precoder for systems without the need to exchange CSI among cells. First, the design of transmit precoding matrices and vectors is tackled by jointly minimizing a sum-MSE of all users subject to individual transmit power constraints at each cell. Based on this formulation, two alternating optimization algorithms named relaxed-constraints based alternating optimization (RAO) and unconstrained alternating optimization with normalization (UAON) are presented, where the RAO relaxes the non-convex constraints involved to convex ones first and then employs an alternating optimization technique to produce the solution, while the UAON is performed by embedding the constraints into the optimization process via a normalization step. Motivated by the techniques aimed at multi-cell time division duplex (TDD) systems~\cite{pilot2011,SCMIMO2013}, next we develop a low complexity precoding scheme for HetNet where the precoder in each cell is designed separately without the need to exchange user data or CSI over the backhaul. By employing block diagonalization (BD) techniques at the node side~\cite{Rui2010}, we derive a two-level precoder by a non-iterative algorithm where different interference thresholds are utilized to control the relative weights associated with the interferences for performance enhancement. Moreover, robust precoding schemes are presented correspondingly with imperfect CSI known at each node. Finally, we present results from numerical experiments for the proposed precoding strategies under different system configurations as well as a comparison study on performance in terms of MSE and bit error rate (BER).

The rest of the paper is organized as follows. The system model for the MIMO downlinks in HetNet systems is described in Section~\ref{sec:sys_model}. In Section~\ref{sec:Sum_Prec}, a new sum-MSE based precoding scheme for HetNet is proposed and two implementation algorithms are elaborated. In Section~\ref{sec:Sep_Prec}, a separate MSE based precoding algorithm is developed for the BS and SCs, respectively, and two-level precoders are derived. Then, robust precoders are designed based on the estimated channel knowledge in Section~\ref{sec:Robust_Prec}. Simulation results for several different system configurations are presented in Section~\ref{sec:shadowing_simulation} to demonstrate the performance of the proposed precoding techniques. Finally, we draw our conclusions in Section~\ref{sec:conclusion}.

\emph{Notations:} We use ${\text {tr}}\{\bf{X}\}$, ${\bf{X}}^T$, ${\bf{X}}^H$ and $\left\| \bf{X} \right\|_F$ to denote the trace, transpose, Hermitian transpose, and Frobenius norm of matrix $\bf{X}$, respectively. The symbol $\left\|{\bf x}\right\|$ denotes the 2-norm of vector $\bf x$, ${\text {diag}}\{{\bf x}\}$ denotes a diagonal matrix with $\bf x$ being its diagonal, and ${\rm bd}\left\{ {{\bf{X}}_1, \ldots ,{\bf{X}}_{K}} \right\}$ denotes a block diagonal matrix with the main diagonal blocks as matrices ${{\bf{X}}_1, \ldots ,{\bf{X}}_{K}}$. The $N\times N$ identity matrix is denoted by ${\bf I}_N$. Furthermore, the expectation of a random variable is denoted by ${\rm E}\{\cdot\}$, and ${\rm vec}\{\cdot\}$ denotes a vector composed of all columns of a matrix in sequence.

\begin{figure}
   \centering
   \includegraphics[scale=0.35]{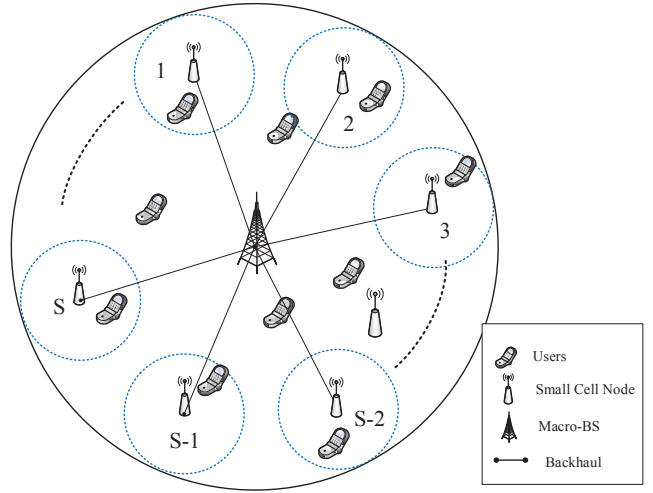}
   \caption{System model for HetNet with SCs deployment.}
   \label{SystemModel}
\end{figure}

\section{System Model}\label{sec:sys_model}
We consider a two-tier network architecture with one cell consisting of one macro BS, which is overlaid with a dense tier of $S$ uniformly distributed SCs as shown in Fig.~\ref{SystemModel}. Assume that the BS and SCs are respectively equipped with $N_{\rm BS}$ and $N_{\rm SC}$ antennas, while each user is dropped uniformly in the cell area and processes $N_{\rm UE}$ antennas. Here, each user is uniformly dropped in the cell area. Based on the maximum reference signal received power (RSRP)~\cite{3gpp36.814}, the users served by the macro BS are assigned to a macro UE (MUE) set, and those served by the SCs are assigned to a small cell UE (SUE) set. Suppose the macro BS serves $K$ MUEs with $K\le N_{\rm BS}$ while $s$-th SC~($s\in \Omega =\{1,2,\dots,S\}$) serves $L_s\le N_{\rm SC}$ SUEs, thus the MUE and $s$-th SUE sets can be denoted by $I=\{1,2,\dots,K\}$ and $J_s=\{1,2,\dots,L_s\}$, respectively.

If the BS and SCs apply linear precoding to serve their UEs during the downlink transmissions, then the received signals at the $i$-th ($i\in I$) MUE and $j$-th ($j\in J_s$) SUE in the $s$-th SC are given by
\begin{equation}\label{receive_i}
\begin{split}
&{\bf{y}}_{{\rm{BS}}}^{(i)} = \sum\limits_{k = 1}^K {\sqrt {{P_{{\rm{BS}}}}} {{\left( {{\bf{H}}_{{\rm{B - M}}}^{(i)}} \right)}^H}{\bf{W}}_{{\rm{BS}}}^{(k)}{\bf{x}}_{{\rm{BS}}}^{(k)}}  + \sum\limits_{s = 1}^S \sum\limits_{l = 1}^{{L_s}} \sqrt {{P_{{\rm{SC}}}}} \\
&~~{{{\left( {{\bf{H}}_{{\rm{S - M}}}^{(s,i)}} \right)}^H}{\bf{W}}_{{\rm{SC}}}^{(s,l)}{\bf{x}}_{{\rm{SC}}}^{(s,l)}}   + {\bf{n}}_{{\rm{BS}}}^{(i)} = {\left( {{\bf{G}}_{{\rm{B - M}}}^{(i)}} \right)^H}{{\bf{W}}_{{\rm{BS}}}}{{\bf{x}}_{{\rm{BS}}}} \\
&~~+ \sum\limits_{s = 1}^S {{{\left( {{\bf{G}}_{{\rm{S - M}}}^{(s,i)}} \right)}^H}{\bf{W}}_{{\rm{SC}}}^{(s)}{\bf{x}}_{{\rm{SC}}}^{(s)}}  + {\bf{n}}_{{\rm{BS}}}^{(i)}
\end{split}
\end{equation}
\begin{equation}\label{receive_j}
\begin{split}
&{\bf{y}}_{{\rm{SC}}}^{(s,j)} = \sum\limits_{k = 1}^K {\sqrt {{P_{{\rm{BS}}}}} {{\left( {{\bf{H}}_{{\rm{B - S}}}^{(s,j)}} \right)}^H}{\bf{W}}_{{\rm{BS}}}^{(k)}{\bf{x}}_{{\rm{BS}}}^{(k)}}  + \sum\limits_{t = 1}^S \sum\limits_{l = 1}^{{L_t}} \sqrt {{P_{{\rm{SC}}}}}\\
&~~{ {{\left( {{\bf{H}}_{{\rm{S - S}}}^{(t,s,j)}} \right)}^H}{\bf{W}}_{{\rm{SC}}}^{(t,l)}{\bf{x}}_{{\rm{SC}}}^{(t,l)}} + {\bf{n}}_{{\rm{SC}}}^{(s,j)} = {\left( {{\bf{G}}_{{\rm{B - S}}}^{(s,j)}} \right)^H}{{\bf{W}}_{{\rm{BS}}}}{{\bf{x}}_{{\rm{BS}}}} \\
&~~ + \sum\limits_{t = 1}^S {{{\left( {{\bf{G}}_{{\rm{S - S}}}^{(t,s,j)}} \right)}^H}{\bf{W}}_{{\rm{SC}}}^{(t)}{\bf{x}}_{{\rm{SC}}}^{(t)}}  + {\bf{n}}_{{\rm{SC}}}^{(s,j)}
\end{split}
\end{equation}
respectively, where ${P_{{\rm{BS}}}}$ and ${P_{{\rm{SC}}}}$ represent the average power at the macro BS and SCs; ${{\bf{H}}_{{\rm{B - M}}}^{(i)}}$ and ${{\bf{H}}_{{\rm{B - S}}}^{(j)}}$ denote the $N_{\rm BS}\times N_{\rm UE}$ channel vectors from the BS to the $i$-th MUE and $j$-th SUE, respectively; ${{\bf{H}}_{{\rm{S - M}}}^{(s,i)}}$ and ${{\bf{H}}_{{\rm{S - S}}}^{(s,j)}}$ denote the $N_{\rm SC}\times N_{\rm UE}$ channel vectors from $s$-th SC to the $i$-th MUE and $j$-th SUE, respectively; ${\bf x}_{{\rm{BS}}}^{(k)}\in {\mathbb C}^{N_{\rm S}\times 1}$ and ${{\bf x}_{{\rm{SC}}}^{(s,j)}}\in {\mathbb C}^{N_{\rm S}\times 1}$ are the complex-valued Gaussian $N_{\rm S}$ transmitted symbol streams from BS to its $k$-th MUE and from $s$-th SC to its own SUE; ${{\bf{W}}_{{\rm{BS}}}^{(k)}}$ and ${{\bf{W}}_{{\rm{SC}}}^{(s,j)}}$ are the $N_{\rm BS}\times N_{\rm S}$ and $N_{\rm SC}\times N_{\rm S}$ precoding matrices, respectively; and ${\bf n}_{{\rm{BS}}}^{(i)}$ and ${\bf n}_{{\rm{SC}}}^{(s,j)}$ are the additive white Gaussian noise vectors with each element of variance $N_0$. Besides, ${{\bf{W}}_{{\rm{BS}}}} = \left[ {{\bf{W}}_{{\rm{BS}}}^{(1)},{\bf{W}}_{{\rm{BS}}}^{(2)}, \ldots, {\bf{W}}_{{\rm{BS}}}^{(K)}} \right]$, ${{\bf{W}}_{{\rm{SC}}}^{(t)}} = \left[ {{\bf{W}}_{{\rm{SC}}}^{(t,1)},{\bf{W}}_{{\rm{SC}}}^{(t,2)}, \ldots, {\bf{W}}_{{\rm{SC}}}^{(t,L_t)}} \right]$, ${{\bf{x}}_{{\rm{BS}}}} = {\left[ {{\bf x}_{{\rm{BS}}}^{(1)};{\bf x}_{{\rm{BS}}}^{(2)}; \ldots; {\bf x}_{{\rm{BS}}}^{(K)}} \right]}$, ${{\bf{x}}_{{\rm{SC}}}^{(t)}} = {\left[ {{\bf x}_{{\rm{SC}}}^{(t,1)};{\bf x}_{{\rm{SC}}}^{(t,2)}; \ldots; {\bf x}_{{\rm{SC}}}^{(t,L_t)}} \right]}$, ${\bf{G}}_{{\rm{B - M}}}^{(i)} = \sqrt {{P_{{\rm{BS}}}}} {\bf{H}}_{{\rm{B - M}}}^{(i)}$, ${\bf{G}}_{{\rm{B - S}}}^{(j)} = \sqrt {{P_{{\rm{BS}}}}} {\bf{H}}_{{\rm{B - S}}}^{(j)}$, ${\bf{G}}_{{\rm{S - M}}}^{(s,i)} = \sqrt {{P_{{\rm{SC}}}}} {\bf{H}}_{{\rm{S - M}}}^{(s,i)}$ and ${\bf{G}}_{{\rm{S - S}}}^{(t,s,j)} = \sqrt {{P_{{\rm{SC}}}}} {\bf{H}}_{{\rm{S - S}}}^{(t,s,j)}$ are defined for analysis simplicity. Moreover, the propagation factor here is defined as the product of a fast fading factor and an amplitude factor that accounts for geometric attenuation and shadow fading. For example, $h_{{\rm{B - M}}}^{(m_1,n_1,i)}$ (the $(m_1,n_1)$-th element of ${{\bf{H}}_{{\rm{B - M}}}^{(i)}}$) and ${h_{{\rm{S - M}}}^{(m_2,n_2,s,i)}}$ (the $(m_2,n_2)$-th element of ${{\bf{H}}_{{\rm{S - M}}}^{(s,i)}}$) in (\ref{receive_i}) assume the form
\begin{equation}\label{channel}
\begin{split}
&h_{{\rm{B - M}}}^{(m_1,n_1,i)} = \sqrt {\beta _{{\rm{B - M}}}^{(i)}} \upsilon _{{\rm{B - M}}}^{(m_1,n_1,i)}\\
&{h_{{\rm{S - M}}}^{(m_2,n_2,s,i)}} = \sqrt {\beta _{{\rm{S - M}}}^{(s,i)}} \upsilon _{{\rm{S - M}}}^{(m_2,n_2,s,i)}
\end{split}
\end{equation}
where $m_1\in \{1,2,\dots,N_{\rm BS}\}$, $m_2\in \{1,2,\dots,N_{\rm SC}\}$, $n_1,~n_2\in \{1,2,\dots,N_{\rm UE}\}$; $\upsilon _{{\rm{B - M}}}^{(m_1,n_1,i)} \sim \mathcal{CN}\left( {0,1} \right)$ and $\upsilon _{{\rm{S - M}}}^{(m_2,n_2,s,i)} \sim \mathcal{CN}\left( {0,1} \right)$ denote the fast fading coefficients; and ${\beta _{{\rm{B - M}}}^{(i)}}$ and ${\beta_{{\rm{S - M}}}^{(s,i)}}$ are the amplitude factors. Because the geometric and shadow fading change slowly over space, ${\beta _{{\rm{B - M}}}^{(i)}}$ and ${\beta_{{\rm{S - M}}}^{(s,i)}}$ are treated as constants with respect to the index of the base station antenna, and we can write
\begin{equation}
\beta _{{\rm{B - M}}}^{(i)} = {\zeta_{\rm{BS}}}{\theta_{\rm{BS}}}\left( {d_{{\rm{B - M}}}^{(i)}} \right),~\beta _{{\rm{S - M}}}^{(s,i)} = {\zeta_{{\rm{SC}}}}{\theta_{\rm{SC}}}\left( {d_{{\rm{S - M}}}^{(s,i)}} \right)
\end{equation}
where ${\zeta_{{\rm{BS}}}}$ and ${\zeta_{{\rm{SC}}}}$ denote the corresponding penetration loss that are independent over all the indices~\cite{3GPP2011}, and functions $\theta_{\rm{BS}}\left({d_{{\rm{B - M}}}^{(i)}}\right)$ and $\theta_{\rm{SC}}\left({d_{{\rm{S - M}}}^{(s,i)}}\right)$ represent the pathloss model at the BS and the SCs, respectively, where the arguments ${d_{{\rm{B - M}}}^{(i)}}$ and ${d_{{\rm{S - M}}}^{(s,i)}}$ are the distance between the BS and the $i$-th MUE and the distance between the $s$-th SC and the $i$-th MUE, respectively. Similar expressions for the propagation factors ${{\bf{H}}_{{\rm{B - S}}}^{(s,j)}}$ and ${{\bf{H}}_{{\rm{S - S}}}^{(t,s,j)}}$ in (\ref{receive_j}) can be obtained. We assume that time division duplex is adopted with channel reciprocity satisfied, i.e., the propagation factor is the same for both forward and reverse links and block fading remains constant for a duration symbols. Hence, exact CSI for the downlinks can be obtained for both BS and SCs.

From (\ref{receive_i}), the signal received at MUEs can be expressed as
\begin{equation}\label{receive_MUE}
{{\bf{y}}_{{\rm{BS}}}} = {\bf{G}}_{{\rm{B - M}}}^H{{\bf{W}}_{{\rm{BS}}}}{{\bf{x}}_{{\rm{BS}}}} + \sum\limits_{s = 1}^S {{{\left( {{\bf{G}}_{{\rm{S - M}}}^{(s)}} \right)}^H}{\bf{W}}_{{\rm{SC}}}^{(s)}{\bf{x}}_{{\rm{SC}}}^{(s)}}  + {{\bf{n}}_{{\rm{BS}}}}
\end{equation}
where ${{\bf{G}}_{{\rm{B - M}}}} = \left[ {{\bf{G}}_{{\rm{B - M}}}^{(1)},{\bf{G}}_{{\rm{B - M}}}^{(2)}, \ldots, {\bf{G}}_{{\rm{B - M}}}^{(K)}} \right]$, ${\bf{G}}_{{\rm{S - M}}}^{(s)} = \left[ {{\bf{G}}_{{\rm{S - M}}}^{(s,1)},{\bf{G}}_{{\rm{S - M}}}^{(s,2)}, \ldots, {\bf{G}}_{{\rm{S - M}}}^{(s,K)}} \right]$, ${{\bf{n}}_{{\rm{BS}}}} = {\left[ {{\bf{n}}_{{\rm{BS}}}^{(1)};{\bf{n}}_{{\rm{BS}}}^{(2)}; \ldots; {\bf{n}}_{{\rm{BS}}}^{(K)}} \right]}$ and ${{\bf{y}}_{{\rm{BS}}}} = {\left[ {{\bf{y}}_{{\rm{BS}}}^{(1)};{\bf{y}}_{{\rm{BS}}}^{(2)}; \ldots; {\bf{y}}_{{\rm{BS}}}^{(K)}} \right]}$. Similarly, from (\ref{receive_j}) the signal received at SUEs of the $s$-th SC is
\begin{equation}\label{receive_SUE}
{\bf{y}}_{{\rm{SC}}}^{(s)} = {\left( {{\bf{G}}_{{\rm{B - S}}}^{(s)}} \right)^H}{{\bf{W}}_{{\rm{BS}}}}{{\bf{x}}_{{\rm{BS}}}} + \sum\limits_{t = 1}^S {{{\left( {{\bf{G}}_{{\rm{S - S}}}^{(t,s)}} \right)}^H}{\bf{W}}_{{\rm{SC}}}^{(t)}{\bf{x}}_{{\rm{SC}}}^{(t)}}  + {\bf{n}}_{{\rm{SC}}}^{(s)}
\end{equation}
where ${{\bf{G}}_{\rm{B - S}}^{(s)}} = \left[ {{\bf{G}}_{{\rm{B - S}}}^{(s,1)},{\bf{G}}_{{\rm{B - S}}}^{(s,2)}, \ldots, {\bf{G}}_{{\rm{B - S}}}^{(s,L_s)}} \right]$, ${\bf{G}}_{{\rm{S - S}}}^{(t,s)} = \left[ {{\bf{G}}_{{\rm{S - S}}}^{(t,s,1)},{\bf{G}}_{{\rm{S - S}}}^{(t,s,2)}, \ldots, {\bf{G}}_{{\rm{S - S}}}^{(t,s,L_s)}} \right]$, ${{\bf{n}}_{{\rm{SC}}}^{(s)}} = {\left[ {{\bf n}_{{\rm{SC}}}^{(s,1)};{\bf n}_{{\rm{SC}}}^{(s,2)}; \ldots; {\bf n}_{{\rm{SC}}}^{(s,L_s)}} \right]}$ and ${{\bf{y}}_{{\rm{SC}}}^{(s)}} = {\left[ {{\bf y}_{{\rm{SC}}}^{(s,1)};{\bf y}_{{\rm{SC}}}^{(s,2)}; \ldots; {\bf y}_{{\rm{SC}}}^{(s,L_s)}} \right]}$.

Assume that the linear receiver is applied at each user, then
\begin{equation}\label{estimated_UE}
{\hat{\bf x}}_{\rm BS}^{(i)} = {\bf R}_{\rm BS}^{(i)}{\bf y}_{\rm BS}^{(i)},~{\hat{\bf x}}_{\rm SC}^{(s,j)} = {\bf R}_{\rm SC}^{(s,j)}{\bf y}_{\rm SC}^{(s,j)},~s\in \Omega
\end{equation}
where ${\bf R}_{\rm BS}^{(i)}\in {\mathbb C}^{N_{\rm S}\times N_{\rm UE}}$ and ${\bf R}_{\rm SC}^{(s,j)}\in {\mathbb C}^{N_{\rm S}\times N_{\rm UE}}$ are the receiving filter matrices of MUE $i$ and SUE $j$ in the $s$-th SC, respectively. For simplicity, in the rest of the paper (\ref{estimated_UE}) is rewritten as
\begin{equation}\label{estimated_UE_simp}
{\hat{\bf x}}_{\rm BS} = {\bf R}_{\rm BS}{\bf y}_{\rm BS},~{\hat{\bf x}}_{\rm SC}^{(s)} = {\bf R}_{\rm SC}^{(s)}{\bf y}_{\rm SC}^{(s)},~s\in \Omega
\end{equation}
where ${\bf R}_{\rm BS}={\rm bd}\left\{ {{\bf{R}}_{{\rm{BS}}}^{(1)}, \ldots ,{\bf{R}}_{{\rm{BS}}}^{(K)}} \right\}$ and ${\bf{R}}_{{\rm{SC}}}^{(s)} = {\rm bd}\left\{ {{\bf{R}}_{{\rm{SC}}}^{(s,1)}, \ldots ,{\bf{R}}_{{\rm{SC}}}^{(s,{L_s})}} \right\}$.

\section{Sum-MSE Minimization Based Precoding in HetNet}\label{sec:Sum_Prec}
In this section, the design of precoding matrices ${{\bf{W}}_{{\rm{BS}}}}$ and ${\bf{W}}_{{\rm{SC}}}^{(s)}$ ($s\in \Omega$) is addressed by minimizing the total MSE (we call it sum-MSE) where each squared error term involves its corresponding receiver matrix ${{\bf{R}}_{{\rm{BS}}}}$ or ${\bf{R}}_{{\rm{SC}}}^{(s)}$ that can be performed by the user. This minimization is carried out subject to average power constraints on ${{\bf{W}}_{{\rm{BS}}}}$ and ${\bf{W}}_{{\rm{SC}}}^{(s)}$ for $s\in \Omega$. Under these circumstances, the precoding design problem can be cast as a constrained optimization problem
\begin{subequations}
    \label{Opt1}
    \begin{align}
    \label{objective1}
    \begin{split}
    &\mathop {\min }\limits_{\begin{array}{*{20}{c}}
    {{{\bf{W}}_{{\rm{BS}}}},{\bf{W}}_{{\rm{SC}}}^{(t)}}\\
    {{{\bf{R}}_{{\rm{BS}}}},{\bf{R}}_{{\rm{SC}}}^{(t)},t \in \Omega }
    \end{array}} {\rm{ E}}\left\{ {{{\left\| {{{\hat{\bf{ x}}}_{{\rm{BS}}}} - {{\bf{x}}_{{\rm{BS}}}}} \right\|}^2} + \sum\limits_{s = 1}^S {{{\left\| {\hat{\bf{ x}}_{{\rm{SC}}}^{(s)} - {\bf{x}}_{{\rm{SC}}}^{(s)}} \right\|}^2}} } \right\}
    \end{split}\\
    \label{constraints1}
    \begin{split}
    &{\rm{subject~to}}~~{\rm {tr}}\left\{ {{\bf{W}}_{{\rm{BS}}}^H{{\bf{W}}_{{\rm{BS}}}}} \right\} = 1
    \end{split}\\
    \label{constraints2}
    \begin{split}
    &~~~~~~~~~~{\rm {tr}}\left\{ {\left({\bf{W}}_{{\rm{SC}}}^{(t)}\right)^H{{\bf{W}}_{{\rm{SC}}}^{(t)}}} \right\} = 1,~{\rm{for}}~t\in \Omega
    \end{split}\\
    \label{constraints3}
    \begin{split}
    &~~~~~~~~~~ {{\bf{R}}_{{\rm{BS}}}} = {\rm bd}\left\{ {{\bf{R}}_{{\rm{BS}}}^{(1)}, \ldots ,{\bf{R}}_{{\rm{BS}}}^{(K)}} \right\}
    \end{split}\\
    \label{constraints4}
    \begin{split}
    &~~~~~~~~~~ {\bf{R}}_{{\rm{SC}}}^{(t)} = {\rm bd}\left\{ {{\bf{R}}_{{\rm{SC}}}^{(t,1)}, \ldots ,{\bf{R}}_{{\rm{SC}}}^{(t,{L_t})}} \right\},~{\rm{for}}~t\in \Omega.
    \end{split}
    \end{align}
\end{subequations}
Let ${\bf W}_{\rm{SC}} = \left[ {\bf{W}}_{{\rm{SC}}}^{(1)};~{\bf{W}}_{{\rm{SC}}}^{(2)}; ~\ldots; ~{\bf{W}}_{{\rm{SC}}}^{(S)} \right]$ and ${\bf R}_{\rm{SC}} = \left[ {\bf R} _{{\rm{SC}}}^{(1)},{\bf R} _{{\rm{SC}}}^{(2)}, \ldots, {\bf R} _{{\rm{SC}}}^{(S)} \right]$, and note that the transmission symbols satisfy ${\rm E}\left\{ {\bf{x}} \right\} = 0$, ${\rm E}\left\{ {{\bf{x}}{{\bf{x}}^H}} \right\} = {\bf{I}}$, and $\left\|{\bf x}\right\|^2={\text {tr}}\{{\bf xx}^H\}$, the objective function in (\ref{objective1}) can be rewritten to make its dependence on ${{\bf{W}}_{{\rm{BS}}}}$ and ${\bf{W}}_{{\rm{SC}}}^{(s)}$ explicit as
\begin{equation}\label{Obj_1}
    f\left( {{{\bf{W}}_{{\rm{BS}}}},{\bf{W}}_{{\rm{SC}}},{{\bf{R}}_{{\rm{BS}}}},{\bf{R}}_{{\rm{SC}}}} \right) = {\rm{MS}}{{\rm{E}}_{{\rm{BS}}}} + {\rm{MS}}{{\rm{E}}_{{\rm{SC}}}}
\end{equation}
where
\begin{equation}\label{Obj_ex1}
    \begin{split}
    &{\rm{MS}}{{\rm{E}}_{{\rm{BS}}}}  \buildrel \Delta \over =  {\rm E}\left\{ {{{\left\| {{{\hat{\bf{ x}}}_{{\rm{BS}}}} - {{\bf{x}}_{{\rm{BS}}}}} \right\|}^2}} \right\}= {\rm tr}\left\{ {{\bf{R}}_{{\rm{BS}}}}\left[ {\bf{G}}_{{\rm{B - M}}}^H{{\bf{W}}_{{\rm{BS}}}}\right.\right.\\
    &\left.{\bf{W}}_{{\rm{BS}}}^H{{\bf{G}}_{{\rm{B - M}}}} + \sum\limits_{s = 1}^S {{\left( {{\bf{G}}_{{\rm{S - M}}}^{(s)}} \right)}^H} {\bf{W}}_{{\rm{SC}}}^{(s)}{{\left( {{\bf{W}}_{{\rm{SC}}}^{(s)}} \right)}^H}{\bf{G}}_{{\rm{S - M}}}^{(s)}\right]{\bf{R}}_{{\rm{BS}}}^H\\
    &\left.- 2{{\bf{R}}_{{\rm{BS}}}}{\bf{G}}_{{\rm{B - M}}}^H{{\bf{W}}_{{\rm{BS}}}} + {{\bf I}_{K{N_{\rm S}}}} + \sigma _0^2{{\bf{R}}_{{\rm{BS}}}}{\bf{R}}_{{\rm{BS}}}^H \right\}
\end{split}
\end{equation}
and
\begin{equation}\label{Obj_ex2}
    \begin{split}
    &{\rm{MS}}{{\rm{E}}_{{\rm{SC}}}} \buildrel \Delta \over = {\rm E}\left\{ {\sum\limits_{s = 1}^S {{{\left\| {\hat{\bf{ x}}_{{\rm{SC}}}^{(s)} - {\bf{x}}_{{\rm{SC}}}^{(s)}} \right\|}^2}} } \right\}= \sum\limits_{s = 1}^S {\rm tr}\left\{ {\bf{R}}_{{\rm{SC}}}^{(s)}\left[ {{\left( {{\bf{G}}_{{\rm{B - S}}}^{(s)}} \right)}^H}\right.\right.\\
    &\left.{{\bf{W}}_{{\rm{BS}}}}{\bf{W}}_{{\rm{BS}}}^H{\bf{G}}_{{\rm{B - S}}}^{(s)} + \sum\limits_{t = 1}^S {{\left( {{\bf{G}}_{{\rm{S - S}}}^{(t,s)}} \right)}^H}{\bf{W}}_{{\rm{SC}}}^{(t)}{{\left( {{\bf{W}}_{{\rm{SC}}}^{(t)}} \right)}^H}{\bf{G}}_{{\rm{S - S}}}^{(t,s)}\right]\\
    &\left.{{\left( {{\bf{R}}_{{\rm{SC}}}^{(s)}} \right)}^H}- 2{\bf{R}}_{{\rm{SC}}}^{(s)}{{\left( {{\bf{G}}_{{\rm{S - S}}}^{(s,s)}} \right)}^H}{\bf{W}}_{{\rm{SC}}}^{(s)} + {{\bf I}_{{L_s}{N_{\rm S}}}} + \sigma _0^2{\bf{R}}_{{\rm{SC}}}^{(s)}{{\left( {{\bf{R}}_{{\rm{SC}}}^{(s)}} \right)}^H} \right\}.
\end{split}
\end{equation}
From (\ref{Obj_ex1}) and (\ref{Obj_ex2}) it follows that the sum-MSE is convex w.r.t ${{\bf{W}}_{{\rm{BS}}}}$ and ${\bf{W}}_{{\rm{SC}}}^{(t)}$; and that it is also convex w.r.t. ${\bf R}_{\rm BS}$ and the matrices in ${\bf R}_{\rm{SC}}$. An essential technical difficulty in dealing with problem (\ref{Opt1}) is that both its objective function and the constraints on average power are nonconvex. In what follows, we propose an alternating convex optimization (ACO) technique which turns out to be well suited for the precoding design problem at hand. Specifically, a significant advantage of using ACO-based techniques is that all sub-problems involved are convex, and fast algorithms for their solutions and reliable software code for implementations are available~\cite{convex2004,cvx2008}. In what follows we present two alternating-optimization based techniques. The first technique partitions the design variables into two subsets such that the objective becomes convex with respect to each subset of variables, and this variable partitioning is done while the constraints on average power are relaxed to their convex counterparts. The second technique carries out unconstrained alternating optimization with respect to the above-mentioned two subsets of design variables alternatively, followed by a simple norm normalization step to satisfy the requirement on average power.

\subsection{Relaxed-constraints based Alternating Optimization (RAO)}\label{constrained}
Here we consider a variant of problem (\ref{Opt1}) by a natural convex relaxation of the nonconvex constraints in (\ref{constraints1}) and (\ref{constraints2}), namely,
\begin{subequations}
    \label{Opt1new}
    \begin{align}
    \label{objective1new}
    \begin{split}
    &\mathop {\min }\limits_{\begin{array}{*{20}{c}}
    {{{\bf{W}}_{{\rm{BS}}}},{\bf{W}}_{{\rm{SC}}}^{(t)}}\\
    {{{\bf{R}}_{{\rm{BS}}}},{\bf{R}}_{{\rm{SC}}}^{(t)},t \in \Omega }
    \end{array}} f\left( {{{\bf{W}}_{{\rm{BS}}}},{\bf{W}}_{{\rm{SC}}},{{\bf{R}}_{{\rm{BS}}}},{\bf{R}}_{{\rm{SC}}}} \right)
    \end{split}\\
    \label{constraints1new}
    \begin{split}
    &~~~{\rm{subject~to}}~~~{\rm {tr}}\left\{ {{\bf{W}}_{{\rm{BS}}}^H{{\bf{W}}_{{\rm{BS}}}}} \right\} \le 1
    \end{split}\\
    \label{constraints2new}
    \begin{split}
    &~~~~~~~~~~~~~~~~~~{\rm {tr}}\left\{ {\left({\bf{W}}_{{\rm{SC}}}^{(t)}\right)^H{{\bf{W}}_{{\rm{SC}}}^{(t)}}} \right\} \le 1,~{\rm{for}}~t\in \Omega
    \end{split}\\
    \label{constraints3new}
    \begin{split}
    &~~~~~~~~~~~~~~~~~~(\ref{constraints3}),~(\ref{constraints4})
    \end{split}
    \end{align}
\end{subequations}
As (\ref{constraints1}) and (\ref{constraints2}) impose conditions on the average power at the BS and SCs, its convex relaxation as seen in (\ref{constraints1new}) and (\ref{constraints2new}) are well justified as it limits the average power at the BS and SCs to be within the given values. As will become transparent shortly, this convex relaxation removes the only obstacle that would otherwise prevent us from applying an ACO-based technique to the precoding problem.

To solve problem (\ref{Opt1new}), we begin by partitioning the design variables into two sets, namely ${{\bf{X}}_1}=\{{{\bf{W}}_{{\rm{BS}}}},{\bf{W}}_{{\rm{SC}}}\}$ and ${{\bf{X}}_2}=\{{{\bf R}_{{\rm{BS}}}},{{\bf R}}_{{\rm{SC}}} \}$. Note that $f\left( {{{\bf{W}}_{{\rm{BS}}}},{\bf{W}}_{{\rm{SC}}},{{\bf{R}}_{{\rm{BS}}}},{\bf{R}}_{{\rm{SC}}}} \right)$ in (\ref{objective1new}) is convex w.r.t. variable set ${\bf X}_1$ while variable set ${\bf X}_2$ is fixed, and that it is also convex w.r.t. ${\bf X}_2$ while ${\bf X}_1$ is fixed. Therefore, it is natural to apply an ACO approach for the solution of (\ref{Opt1new}), which is outlined as follows. With variables in ${\bf X}_1$ fixed, one minimizes convex objective function $f\left( {{\bf{W}}_{\rm{BS}},{\bf{W}}_{\rm{SC}},{\bf{R}}_{\rm{BS}},{\bf{R}}_{\rm{SC}}} \right)$ w.r.t. variables $\{{\bf{R}}_{\rm{BS}},{\bf{R}}_{\rm{SC}} \}$. Clearly this is an unconstrained convex problem because variables $\{{\bf{R} _{\rm{BS}}},{\bf{R}} _{{\rm{SC}}} \}$ are not involved in (\ref{constraints1new}) and (\ref{constraints2new}) and constraints in (\ref{constraints3new}) can be removed by substituting it into the objective function. The solution of the above problem, denoted by ${{\bf{X}}_2^*}=\{{\bf{R}}_{\rm{BS}}^*,{\bf{R}}_{\rm{SC}}^* \}$, are then fixed and one minimizes the convex objective function $f\left( {{\bf{W}}_{\rm{BS}},{\bf{W}}_{\rm{SC}},{\bf{R}}_{\rm{BS}}^*,{\bf{R}}_{\rm{SC}}^*} \right)$ w.r.t. $\{{{\bf{W}}_{{\rm{BS}}}},{\bf{W}}_{{\rm{SC}}}\}$ subject to constraints (\ref{constraints1new}) and (\ref{constraints2new}). Obviously this is a constrained convex problem that can be solved efficiently. Having obtained its solution $\{{{\bf{W}}_{{\rm{BS}}}^*},{\bf{W}}_{{\rm{SC}}}^*\}$, the next round of ACO starts, and the procedure continues until a norm of the variations in both variable sets obtained from the two current consecutive rounds is less than a prescribed tolerance and the most current $\{{{\bf{W}}_{{\rm{BS}}}^*},{\bf{W}}_{{\rm{SC}}}^*,{\bf{R}}_{{\rm{BS}}}^*,{\bf{R}}_{{\rm{SC}}}^*\}$ is taken as the solution of the problem. The technical details of solving the two convex sub-problems now follow.

\subsubsection{With ${{\bf{X}}_1}$ fixed} In this case, ${{\bf{W}}_{{\rm{BS}}}}$ and ${\bf{W}}_{{\rm{SC}}}$ are given and the optimization problem in (\ref{Opt1new}) assumes the form
\begin{subequations}
    \label{Opt5}
    \begin{align}
    \label{Opt5_obj}
    \begin{split}
    \mathop {\min }\limits_{{\bf R} _{{\rm{BS}}},{\bf R} _{{\rm{SC}}}} & f_1\left({{\bf{R}}_{{\rm{BS}}}},{\bf{R}}_{{\rm{SC}}} \right)
    \end{split}\\
    \label{constraints5new}
    \begin{split}
    {\rm{subject~to}}~~&(\ref{constraints3}),~(\ref{constraints4})
    \end{split}
    \end{align}
\end{subequations}
Substituting constraints (\ref{constraints3}) and (\ref{constraints4}) into eq.~(\ref{Obj_1}), it follows (\ref{Obj_1_new1}), where (\ref{Obj1_new1_ext}).
\begin{table*}
\begin{equation}\label{Obj_1_new1}
\begin{split}
&f_1\left({{\bf{R}}_{{\rm{BS}}}},{\bf{R}}_{{\rm{SC}}} \right)=\sum\limits_{i = 1}^K {{\rm tr}\left\{ {{\bf{R}}_{{\rm{BS}}}^{(i)}{\bf{\Psi }}_{{\rm{BS}}}^{(i)}{{\left( {{\bf{R}}_{{\rm{BS}}}^{(i)}} \right)}^H} - 2{\bf{R}}_{{\rm{BS}}}^{(i)}{{\left( {{\bf{G}}_{{\rm{B - M}}}^{(i)}} \right)}^H}{\bf{W}}_{{\rm{BS}}}^{(i)} + {{\bf{I}}_{{N_{\rm{S}}}}} + \sigma _0^2{\bf{R}}_{{\rm{BS}}}^{(i)}{{\left( {{\bf{R}}_{{\rm{BS}}}^{(i)}} \right)}^H}} \right\}}\\
&~+ \sum\limits_{s = 1}^S {\sum\limits_{j = 1}^{{l_s}} {{\rm tr}\left\{ {{\bf{R}}_{{\rm{SC}}}^{(s,j)}{\bf{\Psi }}_{{\rm{SC}}}^{(s,j)}{{\left( {{\bf{R}}_{{\rm{SC}}}^{(s,j)}} \right)}^H} - 2{\bf{R}}_{{\rm{SC}}}^{(s,j)}{{\left( {{\bf{G}}_{{\rm{S - S}}}^{(s,s,j)}} \right)}^H}{\bf{W}}_{{\rm{SC}}}^{(s,j)} + {{\bf{I}}_{{N_{\rm{S}}}}} + \sigma _0^2{\bf{R}}_{{\rm{SC}}}^{(s,j)}{{\left( {{\bf{R}}_{{\rm{SC}}}^{(s,j)}} \right)}^H}} \right\}} }
\end{split}
\end{equation}
\begin{subequations}
    \label{Obj1_new1_ext}
    \begin{align}
    \begin{split}
    {\bf{\Psi }}_{{\rm{BS}}}^{(i)} = {\left( {{\bf{G}}_{{\rm{B - M}}}^{(i)}} \right)^H}{{\bf{W}}_{{\rm{BS}}}}{\bf{W}}_{{\rm{BS}}}^H{\bf{G}}_{{\rm{B - M}}}^{(i)} + \sum\limits_{s = 1}^S {{{\left( {{\bf{G}}_{{\rm{S - M}}}^{(s,i)}} \right)}^H}{\bf{W}}_{{\rm{SC}}}^{(s)}{{\left( {{\bf{W}}_{{\rm{SC}}}^{(s)}} \right)}^H}{\bf{G}}_{{\rm{S - M}}}^{(s,i)}}
    \end{split}\\
    \begin{split}
    {\bf{\Psi }}_{{\rm{SC}}}^{(s,j)} = {\left( {{\bf{G}}_{{\rm{B - S}}}^{(s,j)}} \right)^H}{{\bf{W}}_{{\rm{BS}}}}{\bf{W}}_{{\rm{BS}}}^H{\bf{G}}_{{\rm{B - S}}}^{(s,j)} + \sum\limits_{t = 1}^S {{{\left( {{\bf{G}}_{{\rm{S - S}}}^{(t,s,j)}} \right)}^H}{\bf{W}}_{{\rm{SC}}}^{(t)}{{\left( {{\bf{W}}_{{\rm{SC}}}^{(t)}} \right)}^H}{\bf{G}}_{{\rm{S - S}}}^{(t,s,j)}}
    \end{split}
    \end{align}
\end{subequations}
\end{table*}
Hence, the global minimizer ${{\bf R} _{{\rm{BS}}}}^*$ and ${{\bf R}} _{{\rm{SC}}}^*$ can be found by solving
\begin{equation}\label{Derive_22}
    \begin{split}
    \frac{{\partial f_1\left( {{{\bf{R}}_{{\rm{BS}}}},{{\bf{R}}_{{\rm{SC}}}}} \right)}}{{\partial {\bf{R}}_{{\rm{BS}}}^{(i)}}} = 0,~\frac{{\partial f_1\left( {{{\bf{R}}_{{\rm{BS}}}},{{\bf{R}}_{{\rm{SC}}}}} \right)}}{{\partial {\bf{R}}_{{\rm{SC}}}^{(s,j)}}} = 0
    \end{split}
\end{equation}
which gives
\begin{subequations}
    \label{Derive_23}
    \begin{align}
    \label{item231}
    \begin{split}
    {\bf{R}}_{{\rm{BS}}}^{(i)*} = {\left( {{\bf{W}}_{{\rm{BS}}}^{(i)}} \right)^H}{\bf{G}}_{{\rm{B - M}}}^{(i)}{\left( {{\bf{\Psi }}_{{\rm{BS}}}^{(i)} + \sigma _0^2{{\bf{I}}_{{N_{{\rm{UE}}}}}}} \right)^{ - 1}}
    \end{split}\\
    \label{item232}
    \begin{split}
    {\bf{R}}_{{\rm{SC}}}^{(s,j)*} = {\left( {{\bf{W}}_{{\rm{SC}}}^{(s,j)}} \right)^H}{\bf{G}}_{{\rm{S - S}}}^{(s,s,j)}{\left( {{\bf{\Psi }}_{{\rm{SC}}}^{(s,j)} + \sigma _0^2{{\bf{I}}_{{N_{{\rm{UE}}}}}}} \right)^{ - 1}}
    \end{split}
    \end{align}
\end{subequations}
with $\forall i \in I$, $j \in {J_s}$ and $s \in \Omega$. As we can see, the optimal linear receivers ${{\bf R}_{{\rm{BS}}}^*}$ and ${\bf R}_{{\rm{SC}}}^{(s,j)*}$~($s\in \Omega$) depend on the optimal transmit precoding matrices ${\bf{W}}_{{\rm{BS}}}$ and ${\bf{W}}_{{\rm{SC}}}^{(s,j)}$. In this way, the optimal solution ${{\bf R}_{{\rm{BS}}}^*}$ and ${\bf R}_{{\rm{SC}}}^{(s)*}$~($s\in \Omega$) for problem (\ref{Opt5}) can be easily obtained by (\ref{Derive_23}) based on the assumption of ${\bf X}_1$ being fixed.

\subsubsection{With ${{\bf{X}}_2}$ fixed} In this case, ${{\bf R}_{{\rm{BS}}}}$ and ${\bf R} _{{\rm{SC}}}$ are fixed, and the optimization problem in (\ref{Opt1}) assumes the form
\begin{subequations}
    \label{Opt6}
    \begin{align}
    \label{objective6}
    \begin{split}
    \mathop {\min }\limits_{{{\bf{W}}_{{\rm{BS}}}},{\bf{W}}_{{\rm{SC}}}} &f_2\left( {{{\bf{W}}_{{\rm{BS}}}},{\bf{W}}_{{\rm{SC}}}} \right)
    \end{split}\\
    \label{constraints6}
    \begin{split}
    {\rm{subject~to}}~~&(\ref{constraints1new}),~(\ref{constraints2new}).
    \end{split}
    \end{align}
\end{subequations}
With $\lambda_0$ and $\lambda_s$~($s\in \Omega$) as the Lagrange multipliers associated with the power constraints, the Lagrangian of problem (19) is given by~\cite{convex2004}
\begin{equation}\label{Derive_4}
\begin{split}
&L\left( {{{\bf{W}}_{{\rm{BS}}}},{{\bf{W}}_{{\rm{SC}}}},{\bf \lambda}} \right) = f_2\left( {{{\bf{W}}_{{\rm{BS}}}},{\bf{W}}_{{\rm{SC}}}} \right) + {\lambda _0}\left[ {\rm{tr}}\left\{ {{\bf{W}}_{{\rm{BS}}}^H{{\bf{W}}_{{\rm{BS}}}}} \right\} \right.\\
&~~~~~~~~~~\left.- 1 \right]+ \sum\limits_{s = 1}^S {{\lambda _s}\left[ {{\rm{tr}}\left\{ {{{\left( {{\bf{W}}_{{\rm{SC}}}^{(s)}} \right)}^H}{\bf{W}}_{{\rm{SC}}}^{(s)}} \right\} - 1} \right]}
\end{split}
\end{equation}
where for notation simplicity we have defined ${\bf {\lambda}} = {\left[ {\lambda_0},{\lambda_1}, \ldots, {\lambda_S} \right]^T}$. Given ${\bf R}_{\rm{BS}}$, ${\bf R}_{\rm{SC}}$, $\lambda_0$ and $\lambda_s$~($s\in \Omega$), the Lagrangian in (\ref{Derive_4}) is minimized if and only if
\begin{equation}
\frac{{\partial L\left( {{{\bf{W}}_{{\rm{BS}}}},{{\bf{W}}_{{\rm{SC}}}},{\bf{\lambda }}} \right)}}{{\partial {{\bf{W}}_{{\rm{BS}}}}}} = 0,~\frac{{\partial L\left( {{{\bf{W}}_{{\rm{BS}}}},{{\bf{W}}_{{\rm{SC}}}},{\bf{\lambda }}} \right)}}{{\partial {\bf{W}}_{{\rm{SC}}}^{(s)}}} = 0,~s \in \Omega
\end{equation}
i.e.,
\begin{subequations}
    \label{Opt_w}
    \begin{align}
    \label{W}
    \begin{split}
    {\bf{W}}_{{\rm{BS}}}^* = {\left( {{{\bf{\Phi }}_{{\rm{BS}}}} + {\lambda _0}{{\bf{I}}_{{N_{{\rm{BS}}}}}}} \right)^{ - 1}}{{\bf{G}}_{{\rm{B - M}}}}{\bf{R}}_{{\rm{BS}}}^H
    \end{split}\\
    \label{ws}
    \begin{split}
    {\bf{W}}_{{\rm{SC}}}^{(s)*} = {\left( {{\bf{\Phi }}_{{\rm{SC}}}^{(s)} + {\lambda _s}{{\bf{I}}_{{N_{{\rm{SC}}}}}}} \right)^{ - 1}}{\bf{G}}_{{\rm{S - S}}}^{(s,s)}{\left( {{\bf{R}}_{{\rm{SC}}}^{(s)}} \right)^H}.
    \end{split}
    \end{align}
\end{subequations}
where (\ref{Opt_w_ext}).
\begin{table*}
\begin{subequations}
    \label{Opt_w_ext}
    \begin{align}
    \begin{split}
    {{\bf{\Phi }}_{{\rm{BS}}}} = {{\bf{G}}_{{\rm{B - M}}}}{\bf{R}}_{{\rm{BS}}}^H{{\bf{R}}_{{\rm{BS}}}}{\bf{G}}_{{\rm{B - M}}}^H + \sum\limits_{s = 1}^S {{\bf{G}}_{{\rm{B - S}}}^{(s)}{{\left( {{\bf{R}}_{{\rm{SC}}}^{(s)}} \right)}^H}{\bf{R}}_{{\rm{SC}}}^{(s)}{{\left( {{\bf{G}}_{{\rm{B - S}}}^{(s)}} \right)}^H}}
    \end{split}\\
    \begin{split}
    {\bf{\Phi }}_{{\rm{SC}}}^{(s)} = {\bf{G}}_{{\rm{S - M}}}^{(s)}{\bf{R}}_{{\rm{BS}}}^H{{\bf{R}}_{{\rm{BS}}}}{\left( {{\bf{G}}_{{\rm{S - M}}}^{(s)}} \right)^H} + \sum\limits_{t = 1}^S {{\bf{G}}_{{\rm{S - S}}}^{(s,t)}{{\left( {{\bf{R}}_{{\rm{SC}}}^{(t)}} \right)}^H}{\bf{R}}_{{\rm{SC}}}^{(t)}{{\left( {{\bf{G}}_{{\rm{S - S}}}^{(s,t)}} \right)}^H}} ,~s\in \Omega.
    \end{split}
    \end{align}
\end{subequations}
\end{table*}
To obtain non-negative multipliers $\lambda_0$ and $\lambda_s$ ($s\in \Omega$) in the above equations, we substitute (\ref{Opt_w}) into (\ref{Derive_4}) and write $L\left( {\bf{\lambda}} \right)=L\left( {{{\bf{W}}_{{\rm{BS}}}^*},{\bf{W}}_{{\rm{SC}}}^*,{\bf{\lambda}}} \right)$. From the complementarity equalities in the Karush-Kuhn-Tucker (KKT) conditions for (\ref{Opt6}), namely
\begin{subequations}
    \label{Complement}
    \begin{align}
    \label{lambda0}
    \begin{split}
    {\lambda _0}\left( {{\rm {tr}}\left\{ {{\bf{W}}_{{\rm{BS}}}^H{{\bf{W}}_{{\rm{BS}}}}} \right\} - 1} \right)=0
    \end{split}\\
    \label{lambdas}
    \begin{split}
    {\lambda _s}\left[{{\rm {tr}}\left\{{{\left( {{\bf{W}}_{{\rm{SC}}}^{(s)}} \right)}^H}{\bf{W}}_{{\rm{SC}}}^{(s)}\right\} - 1} \right]=0,~s\in \Omega
    \end{split}
    \end{align}
\end{subequations}
we see that the optimal Lagrange multipliers are either positive such that the equality constraints in (\ref{constraints1}) hold or zeros such that the constraints in (\ref{constraints1new}) hold strictly. Recalling that the equality constraints in (\ref{constraints1}) are relaxed to the convex inequalities, we first assume that all the multipliers are greater than zero so that the equalities in constraints (\ref{constraints1new}) hold. This is the same as stating that taking partial derivative of $L$ w.r.t. $\lambda_0$ and $\lambda_s$~($s\in \Omega$) yields zero values.

Given that ${{\bf{\Phi}}_{\rm{BS}}}$ can be factorized in the form ${\bf{S}}_{{\rm{BS}}}^H{{\bf{D}}_{{\rm{BS}}}}{{\bf{S}}_{{\rm{BS}}}}$ where ${\bf{S}}_{{\rm{BS}}}^H{{\bf{S}}_{{\rm{BS}}}} = {{\bf{I}}_{N_{\rm {BS}}}}$ and ${{\bf{D}}_{\rm{BS}}} = {\rm {diag}}\left\{ {{d_{{\rm{BS}}}^{\left( 1 \right)}},{d_{{\rm{BS}}}^{\left( 2 \right)}}, \ldots, {d_{{\rm{BS}}}^{\left( {N_{\rm {BS}}} \right)}}} \right\}$, and that each ${{\bf{\Phi}}_{{\rm{SC}}}^{(s)}}$ can be expressed as ${\left( {{\bf{S}}_{{\rm{SC}}}^{(s)}} \right)^H}{\bf{D}}_{{\rm{SC}}}^{(s)}{\bf{S}}_{{\rm{SC}}}^{(s)}$, with ${\left( {{\bf{S}}_{{\rm{SC}}}^{(s)}} \right)^H}{{\bf{S}}_{{\rm{SC}}}^{(s)}} = {\bf{I}}_{N_{\rm {SC}}}$ and ${\bf{D}}_{{\rm{SC}}}^{(s)} = {\rm {diag}}\left\{ {d_{{\rm{SC}}}^{(s,1)}, d_{{\rm{SC}}}^{(s,2)}, \ldots ,d_{{\rm{SC}}}^{(s,{N_{{\rm{SC}}}})}} \right\}$, the Lagrangian $L\left( {\bf{\lambda}} \right)$ can be simplified to an explicit expression in terms of $\lambda_0,~\lambda_1,~\ldots,~\lambda_S$, see (\ref{L_Lambda_simp2}) in Appendix \ref{Appendix_B}. Differentiating $L\left( {\bf{\lambda}} \right)$ in (\ref{L_Lambda_simp2}) w.r.t. $\lambda_0$ and $\lambda_s$~($s\in \Omega$) and setting the results to zero yield
\begin{subequations}
    \label{Derive_8}
    \begin{align}
    \label{item82}
    \begin{split}
     &\frac{\partial L\left( {\bf{\lambda}} \right)}{\partial {\lambda _0}}= \sum\limits_{n = 1}^{N_{\rm{BS}}} {\frac{a_{\rm{BS}}^{(n)}}{\left( {d_{\rm {BS}}^{(n)}} + {\lambda _0} \right)^2}}  - 1 \buildrel \Delta \over = {\chi _0}\left( {\lambda _0} \right)= 0
    \end{split}\\
    \label{item83}
    \begin{split}
     &\frac{\partial L\left( {\bf{\lambda}} \right)}{\partial {\lambda _s}} = \sum\limits_{n = 1}^{N_{\rm{SC}}} {\frac{a_{\rm{SC}}^{(s,n)}}{\left( {d_{\rm {SC}}^{(s,n)}} + {\lambda _s} \right)^2}}  - 1 \buildrel \Delta \over = {\chi _s}\left( {\lambda _s} \right)= 0,~s \in \Omega
    \end{split}
    \end{align}
\end{subequations}
where ${{\bf{A}}_{{\rm{BS}}}} = {\bf{S}}_{{\rm{BS}}}^H{{\bf{G}}_{{\rm{B - M}}}}{\bf{R}}_{{\rm{BS}}}^H{{\bf{R}}_{{\rm{BS}}}}{\bf{G}}_{{\rm{B - M}}}^H{{\bf{S}}_{{\rm{BS}}}}$ is defined with its $(n,n)$-th entry denoted as $a_{\rm{BS}}^{(n)}$, and ${\bf{A}}_{{\rm{SC}}}^{(s)} = {\left( {{\bf{S}}_{{\rm{SC}}}^{(s)}} \right)^H}{\bf{G}}_{{\rm{S - S}}}^{(s,s)}{\left( {{\bf{R}}_{{\rm{SC}}}^{(s)}} \right)^H}{\bf{R}}_{{\rm{SC}}}^{(s)}{\left( {{\bf{G}}_{{\rm{S - S}}}^{(s,s)}} \right)^H}{\bf{S}}_{{\rm{SC}}}^{(s)}$ is defined with its $(n,n)$-th entry denoted as $a_{\rm{SC}}^{(s,n)}$. Based on the equations in (\ref{Derive_8}), we propose a bisection search algorithm to compute the numerical values of the optimal Lagrange multipliers $\lambda_s$ ($s\in \{0,\Omega\}$). The reader is referred to Algorithm 1 for a step-by-step description of the search method.

By substituting the optimal ${\lambda _s^*}~(s\in \{0,\Omega\})$ obtained into (\ref{Opt_w}), the optimal ${{\bf{W}}_{{\rm{BS}}}^*},{\bf{W}}_{{\rm{SC}}}^{(s)*}$ $(s\in \Omega)$ can be calculated according to (\ref{Opt_w}), where ${\bf X}_2$ is assumed to be fixed. As the alternating convex minimization continues, the objective function in (\ref{objective1new}) monotonically decreases that ensures the algorithm’s convergence because the objective function is nonnegative hence it is bounded from below. In practice, the alternating minimization is run sufficient number of times so as to reach a steady-state hence practically optimal design. The reader is referred to Algorithm 2 for a step-by-step summary of the proposed method.

\subsection{Unconstrained Alternating Optimization with Normalization (UAON)}\label{unconstrained}
As will be demonstrated later in Section~\ref{sec:shadowing_simulation}, the RAO algorithm described above offers superior performance, but at the cost of considerable complexity. Below we present an alternative solution for the sum-MSE problem based on unconstrained alternating convex optimization combined with a simple normalization step. More precisely, by relaxing the equality constraints in (\ref{constraints1}) to constraints on average power which are in turn satisfied by normalizing the ${{\bf{\bar W}}_{{\rm{BS}}}}$ and ${\bf{\bar W}}_{{\rm{SC}}}^{(s)}~(s\in \Omega)$ obtained by minimizing the objective function without constraints, optimal precoding can be achieved quickly with reduced complexity relative to that of the RAO algorithm. The technical details that materialize this approach are given as follows.

\subsubsection{With ${{\bf{X}}_1}$ fixed} The optimal ${\bf R} _{{\rm{BS}}}^*$ and ${\bf R} _{{\rm{SC}}}^{(s)*}$~($s\in \Omega$) can be acquired in the same way as the constrained alternating optimization, which results in (\ref{Derive_23}).

\subsubsection{With ${{\bf{X}}_2}$ fixed} Given ${\bf R} _{{\rm{BS}}}$ and ${\bf R} _{{\rm{SC}}}^{(s)}$~($s\in \Omega$), the optimization problem becomes
\begin{equation}\label{Opt4}
    \begin{split}
    \mathop {\min }\limits_{{{\bf{W}}_{{\rm{BS}}}},{\bf{W}}_{{\rm{SC}}}} &f_2\left( {{{\bf{W}}_{{\rm{BS}}}},{\bf{W}}_{{\rm{SC}}}} \right)
    \end{split}
\end{equation}
where no constraints are imposed. Consequently, the global minimizer ${{\bf{W}}_{{\rm{BS}}}^*},{\bf{W}}_{{\rm{SC}}}^{(s)*}~(s\in \Omega)$ are obtained by solving~\cite{convex2004}
\begin{equation}
    \label{Derive_24}
    \frac{{\partial f_2\left( {{{\bf{W}}_{{\rm{BS}}}},{\bf{W}}_{{\rm{SC}}}} \right)}}{{\partial {{\bf{W}}_{{\rm{BS}}}}}} = 0,~\frac{{\partial f_2\left( {{{\bf{W}}_{{\rm{BS}}}},{\bf{W}}_{{\rm{SC}}}} \right)}}{{\partial {\bf{W}}_{{\rm{SC}}}^{(s)}}}= 0,~s \in \Omega
\end{equation}
which yield
\begin{subequations}
    \label{Derive_25}
    \begin{align}
    \label{W2}
    \begin{split}
    {{{\bf{\bar W}}}_{{\rm{BS}}}} = {\bf{\Phi }}_{{\rm{BS}}}^{ - 1}{{\bf{G}}_{{\rm{B - M}}}}{\bf{R}}_{{\rm{BS}}}^H
    \end{split}\\
    \label{ws2}
    \begin{split}
    {\bf{\bar W}}_{{\rm{SC}}}^{(s)} = {\left( {{\bf{\Phi }}_{{\rm{SC}}}^{(s)}} \right)^{ - 1}}{\bf{G}}_{{\rm{S - S}}}^{(s,s)}{\left( {{\bf{R}}_{{\rm{SC}}}^{(s)}} \right)^H},~s \in \Omega.
    \end{split}
    \end{align}
\end{subequations}
Then, the normalized optimal solutions are expressed as
\begin{subequations}
    \label{Derive_26}
    \begin{align}
    \label{item231}
    \begin{split}
    {{\bf{W}}_{{\rm{BS}}}^*} = \frac{{{{{\bf{\bar W}}}_{{\rm{BS}}}}}}{\sqrt{{\rm {tr}}\left\{ {{\bf{\bar W}}_{{\rm{BS}}}^H{{{\bf{\bar W}}}_{{\rm{BS}}}}} \right\}}}
    \end{split}\\
    \label{item232}
    \begin{split}
    {\bf{W}}_{{\rm{SC}}}^{(s)*} = \frac{{{\bf{\bar W}}_{{\rm{SC}}}^{(s)}}}{\sqrt{{{\left( {{\bf{\bar W}}_{{\rm{SC}}}^{(s)}} \right)}^H}{\bf{\bar W}}_{{\rm{SC}}}^{(s)}}},~s \in \Omega.
    \end{split}
    \end{align}
\end{subequations}

\vspace*{1.6em}
\begin{table}[htbp]
\vspace{-1.2em}
\label{tab:1}       
\centering
\begin{tabular}[t]{p{250pt}}
\hline\noalign{\smallskip}
\textbf{Algorithm 1:} Bisection search algorithm \\
\noalign{\smallskip}\hline\noalign{\smallskip}
\textbf{Decide search region:} Calculate ${\chi _s}\left( 0 \right)$ and decide the search region. If ${\chi _s}\left( 0 \right)>0$, find a $\tilde\lambda_s$ satisfying ${\chi _s}\left( \tilde\lambda_s \right)\ge0$ and then go to the initialization step. Otherwise, output $\lambda_s^* = 0$ as the solution.\\
\textbf{Initialize:} Set ${\lambda _{s,{\rm{min}}}}=0$, ${\lambda _{s,{\rm{max}}}}=\tilde\lambda_s$ and a tolerance $\varepsilon$.\\
\textbf{Repeat:}
\begin{enumerate}
    \item Set $\lambda_s ={{\left( {{\lambda _{s,{\rm{min}}}} + {\lambda _{s,{\rm{max}}}}} \right)} \mathord{\left/
 {\vphantom {{\left( {{\lambda _{s,{\rm{min}}}} + {\lambda _{s,{\rm{max}}}}} \right)} 2}} \right.
 \kern-\nulldelimiterspace} 2}$;
    \item Calculate ${\chi _s}\left( \lambda_s \right)$;
    \item Update the search region: If ${\chi _s}\left( \lambda_s \right)\ge 0$, set lower bound to ${\lambda _{s,{\rm {min}}}}=\lambda_s$. If ${\chi _s}\left( \lambda_s \right) < 0$, set upper bound to ${\lambda _{s,{\rm {max}}}}=\lambda_s$.
\end{enumerate}
\textbf{Until:} ${\chi _s}\left( {\lambda _{s,{\rm {min}}}} \right)-{\chi _s}\left( {\lambda _{s,{\rm {max}}}} \right) < \varepsilon$ (search error is less than tolerance).\\
\textbf{Output:} Output $\lambda_s^* ={{\left( {{\lambda _{s,{\rm{min}}}} + {\lambda _{s,{\rm{max}}}}} \right)} \mathord{\left/
 {\vphantom {{\left( {{\lambda _{s,{\rm{min}}}} + {\lambda _{s,{\rm{max}}}}} \right)} 2}} \right.
 \kern-\nulldelimiterspace} 2}$ as the solution.\\
 \hline
\end{tabular}
\end{table}

\vspace*{1.6em}
\begin{table}[htbp]
\vspace{-1em}
\label{tab:2}       
\centering
\begin{tabular}[t]{p{250pt}}
\hline\noalign{\smallskip}
\textbf{Algorithm 2:} RAO \\
\noalign{\smallskip}\hline\noalign{\smallskip}
\textbf{Initialize:} Input initial ${{\bf{R}}_{{\rm{BS}}}^{(0)}}$, ${\bf{R}}_{{\rm{SC}}}^{(s)(0)}$~($s\in \Omega$) and a maximum number of iterations $N_{\rm{iter}}$. Set $k=1$.\\
\textbf{Repeat:}
\begin{enumerate}
    \item Calculate optimal ${\lambda _s^*}~(s\in \{0,\Omega\})$ in (\ref{Derive_8}) by Algorithm 1;
    \item Calculate ${{\bf{W}}_{{\rm{BS}}}^{(k)}}$ and ${\bf{W}}_{{\rm{SC}}}^{(s)(k)}~(s\in \Omega)$ using (\ref{Opt_w});
    \item Calculate optimal ${{\bf{R}}_{{\rm{BS}}}^{(k)}}$ and ${\bf{R}}_{{\rm{SC}}}^{(s)(k)}$~($s\in \Omega$) by substituting the ${{\bf{W}}_{{\rm{BS}}}^{(k)}}$ and ${\bf{W}}_{{\rm{SC}}}^{(s)(k)}~(s\in \Omega)$ obtained in step 1) into (\ref{Derive_23});
    \item Set $k = k+1$.
\end{enumerate}
\textbf{Until:} $k=N_{\rm{iter}}$.\\
\textbf{Output:} Output ${{\bf R} _{{\rm{BS}}}^{(N_{\rm{iter}})}}$, ${\bf R} _{{\rm{SC}}}^{(s)(N_{\rm{iter}})}$~($s\in \Omega$), ${{\bf{W}}_{{\rm{BS}}}^{(N_{\rm{iter}})}}$ and ${\bf{W}}_{{\rm{SC}}}^{(s)(N_{\rm{iter}})}~(s\in \Omega)$ as the solution.\\
 \hline
\end{tabular}
\end{table}

\vspace*{1.6em}
\begin{table}[htbp]
\vspace{-1em}
\label{tab:3}       
\centering
\begin{tabular}[t]{p{250pt}}
\hline\noalign{\smallskip}
\textbf{Algorithm 3:} UAON\\
\noalign{\smallskip}\hline\noalign{\smallskip}
\textbf{Initialize:} Set initial ${{\bf R} _{{\rm{BS}}}^{(0)}}$, ${\bf R}_{{\rm{SC}}}^{(s)(0)}$~($s\in \Omega$) and a maximum number of iterations $N_{\rm{iter}}$. Set $k=1$.\\
\textbf{Repeat:}
\begin{enumerate}
    \item Calculate ${{\bf{\bar W}}_{{\rm{BS}}}^{(k)}}$ and ${\bf{\bar W}}_{{\rm{SC}}}^{(s)(k)}~(s\in \Omega)$ using (\ref{Derive_25}), and normalize them using (\ref{Derive_26}) to obtain ${{\bf{W}}_{{\rm{BS}}}^{(k)}},{\bf{W}}_{{\rm{SC}}}^{(s)(k)}~(s\in \Omega)$;
    \item Calculate optimal ${{\bf R} _{{\rm{BS}}}^{(k)}}$ and ${\bf R} _s^{{\rm{SC}}(k)}$~($s\in \Omega$) by substituting the ${{\bf{W}}_{{\rm{BS}}}^{(k)}}$ and ${\bf{W}}_{{\rm{SC}}}^{(s)(k)}~(s\in \Omega)$ obtained in step 1) into (\ref{Derive_23});
    \item Set $k = k+1$.
\end{enumerate}
\textbf{Until:} $k=N_{\rm{iter}}$.\\
\textbf{Output:} Output ${{\bf R}_{{\rm{BS}}}^{(N_{\rm{iter}})}}$, ${\bf R}_{{\rm{SC}}}^{(s)(N_{\rm{iter}})}$~($s\in \Omega$), ${{\bf{W}}_{{\rm{BS}}}^{(N_{\rm{iter}})}}$ and ${\bf{W}}_{{\rm{SC}}}^{(s)(N_{\rm{iter}})}~(s\in \Omega)$ as the solution.\\
 \hline
\end{tabular}
\end{table}

The reader is referred to Algorithm 3 for a step-by-step summary of the proposed method. We remark that although UAON is much simpler than RAO, the optimal precoder based on UAON still requires the knowledge about the channels from the nodes to both MUEs and SUEs.

\section{Separate MSE Minimization Based Two-level Precoding in HetNet}\label{sec:Sep_Prec}
In Section \ref{sec:Sum_Prec}, the precoders at the BS and all the SCs are jointly designed by minimizing the sum-MSE. However, due to the non-convexity of the objective functions and constraints, no non-iterative algorithms are available for the precoder designs and intensive computation is required. In this section, a simplified solution procedure is derived based on separate MSE minimization where block diagonalization techniques act in the first-level and the second-level precoders at each node are designed separately. As shown in what follows, the separate treatment of individual precoders leads to a non-iterative algorithm.

\subsection{MSE Minimization at the BS}\label{sec:BS}
In order to determine the precoding matrix ${\bf{W}}_{{\rm{BS}}}$ at the BS, the signal and interference associated with the BS are taken into account in a way similar to~\cite{pilot2011}. This leads to
\begin{subequations}
    \label{Opt2}
    \begin{align}
    \label{objective2}
    \begin{split}
    \mathop {\min }\limits_{{{\bf{W}}_{{\rm{BS}}}},{{\bf{R}}_{{\rm{BS}}}}} & {\rm{ E}}\left\{ {{{\left\| {{{\hat{\bf{ x}}}_{{\rm{BS}}}} - {{\bf{x}}_{{\rm{BS}}}}} \right\|}^2}} \right\}
    \end{split}\\
    \label{constraints21}
    \begin{split}
    {\rm{subject~to}}~~&{\left\| {{\bf{G}}_{{\rm{B - S}}}^H{{\bf{W}}_{{\rm{BS}}}}{{\bf{x}}_{{\rm{BS}}}}} \right\|^2} \le \gamma_{\rm{BS}}
    \end{split}\\
    \label{constraints22}
    \begin{split}
    &{\rm {tr}}\left\{ {{\bf{W}}_{{\rm{BS}}}^H{{\bf{W}}_{{\rm{BS}}}}} \right\} \le 1
    \end{split}
    \end{align}
\end{subequations}
where $\gamma_{\rm{BS}} > 0$ is a threshold parameter set to control the relative interference involved. The item in the objective in (\ref{objective2}) is the sum of squares of errors seen by the MUEs assuming no interferences included, i.e., ${{\bf{y}}_{{\rm{BS}}}} = {\bf{G}}_{{\rm{B - M}}}^H{{\bf{W}}_{{\rm{BS}}}}{{\bf{x}}_{{\rm{BS}}}} + {{\bf{n}}_{{\rm{BS}}}}$; and the item in (\ref{constraints21}) is the sum of squares of the interference seen by the SUEs. By tuning $\gamma_{\rm{BS}}$, the BS trades off the beamforming gains for its target MUEs against interference reduction to the neighboring SUEs. We stress that the objective function is not jointly convex with respect to all design variables, but that the induced interference constraint in (\ref{constraints21}) and the average power constraint in (\ref{constraints22}) are convex. Certainly, similar to the RAO proposed in Section \ref{sec:Sum_Prec}, an iterative algorithm could provide an optimal solution for (\ref{Opt2}). To obtain a non-iterative algorithm, we further simplify (\ref{Opt2}) by employing BD technique at the BS side as a first-level precoder. Thereby, all the inter-MUE interferences are eliminated and each MUE perceives an interference-free MIMO channel, which means that problem in (\ref{Opt2}) can be divided into $K$ independent sub-problems of the form
\begin{subequations}
    \label{Opt2NEW}
    \begin{align}
    \label{objective2NEW}
    \begin{split}
    \mathop {\min }\limits_{{\bf{W}}_{{\rm{BS}}}^{(i)},{\bf{R}}_{{\rm{BS}}}^{(i)}}~& {\rm{E}}\left\{ {{{\left\| {{\hat{\bf x}}_{{\rm{BS}}}^{(i)} - {\bf{x}}_{{\rm{BS}}}^{(i)}} \right\|}^2}} \right\}
    \end{split}\\
    \label{constraints21NEW}
    \begin{split}
    {\rm{subject~to}}~~&{\left\| {{\bf{G}}_{{\rm{B - S}}}^H{\bf{W}}_{{\rm{BS}}}^{(i)}{\bf{x}}_{{\rm{BS}}}^{(i)}} \right\|^2} \le \gamma _{{\rm{BS}}}^{(i)}
    \end{split}\\
    \label{constraints22NEW}
    \begin{split}
    &{\rm{tr}}\left\{ {{{\left( {{\bf{W}}_{{\rm{BS}}}^{(i)}} \right)}^H}{\bf{W}}_{{\rm{BS}}}^{(i)}} \right\} \le \alpha _{{\rm{BS}}}^{(i)}
    \end{split}\\
    \label{constraints23NEW}
    \begin{split}
    &{\left( {{\bf{\bar G}}_{{\rm{B - M}}}^{(i)}} \right)^H}{\bf{W}}_{{\rm{BS}}}^{(i)} = {\bf 0}
    \end{split}
    \end{align}
\end{subequations}
where ${\bf{\bar G}}_{{\rm{B - M}}}^{(i)} = \left[ {{\bf{G}}_{{\rm{B - M}}}^{(1)}, \ldots ,{\bf{G}}_{{\rm{B - M}}}^{(i - 1)},{\bf{G}}_{{\rm{B - M}}}^{(i + 1)}, \ldots {\bf{G}}_{{\rm{B - M}}}^{(K)}} \right]$, $i\in I$, $\sum\limits_{i = 1}^K {\gamma _{{\rm{BS}}}^{(i)}}= \gamma_{{\rm{BS}}}$ and $\sum\limits_{i = 1}^K {\alpha _{{\rm{BS}}}^{(i)}} = 1$ with $\gamma _{{\rm{BS}}}^{(i)}>0$ and $\alpha _{{\rm{BS}}}^{(i)}>0$. Here, the BD constraint of (\ref{constraints23NEW}) is imposed to eliminate all inter-MUE interferences. By applying the SVD, we have ${\bar{\bf G}}_{{\rm{B - M}}}^{(i)} = {\bf{U}}_{{\rm{BS}}}^{(i)}{\bf{Z}}_{{\rm{BS}}}^{(i)}{\left[ {{\bf{V}}_{{\rm{BS}}}^{(i,1)}~{\bf{V}}_{{\rm{BS}}}^{(i,0)}} \right]^H}$, where ${\bf{Z}}_{{\rm{BS}}}^{(i)}$ is the diagonal matrix with non-negative singular values as its diagonal elements, ${\bf{V}}_{{\rm{BS}}}^{(i,1)}$ contains the singular vectors corresponding to the nonzero singular values and ${\bf{V}}_{{\rm{BS}}}^{(i,0)}$ consists of vectors corresponding to the zero singular values. Hence, ${\bf{V}}_{{\rm{BS}}}^{(i,0)}$ is an orthogonal basis for the null space of ${\bar{\bf G}}_{{\rm{B - M}}}^{(i)}$. For simplicity, we suppose that ${\bf{W}}_{{\rm{BS}}}^{(i)} = {\bf{W}}_{{\rm{BS}},1}^{(i)}{\bf{W}}_{{\rm{BS}},2}^{(i)}$ for $\forall i \in I$ with ${\bf{W}}_{{\rm{BS}},1}^{(i)}={\bf{V}}_{{\rm{BS}}}^{(i,0)}$ to satisfy the BD constraint of (\ref{constraints23NEW}). In this way, we transform our focus from the design of ${\bf{W}}_{\rm{BS}}^{(i)}$ to that of ${\bf{W}}_{{\rm{BS}},2}^{(i)}$.

Similar to the RAO algorithm, suppose that ${\bf{W}}_{\rm{BS}}^{(i)}$ are fixed, then the optimal ${\bf{R}}_{{\rm{BS}}}^{(i)}$~($i \in I$) can be expressed as (\ref{Opt2_sepRBS}).
\begin{table*}
\begin{equation}\label{Opt2_sepRBS}
    \begin{split}
    {\bf{R}}_{{\rm{BS}}}^{(i)*} = {\left( {{\bf{W}}_{{\rm{BS}}}^{(i)}} \right)^H}{\bf{G}}_{{\rm{B - M}}}^{(i)}{\left[ {{{\left( {{\bf{G}}_{{\rm{B - M}}}^{(i)}} \right)}^H}{{\bf{W}}_{{\rm{BS}}}^{(i)}}\left({\bf{W}}_{{\rm{BS}}}^{(i)}\right)^H{\bf{G}}_{{\rm{B - M}}}^{(i)} + \sigma _0^2{{\bf{I}}_{{N_{{\rm{UE}}}}}}} \right]^{ - 1}}.
    \end{split}
\end{equation}
\begin{equation}\label{Opt2_Obj2}
    \begin{split}
    {\rm{MSE}}_{{\rm{BS}}}^{(i)} = {N_{\rm S}}-{\rm{tr}}\left\{ {{{\left[ {\sigma _0^2{{\left( {{{\left( {{\bf{G}}_{{\rm{B - M}}}^{(i)}} \right)}^H}{{\bf{W}}_{{\rm{BS}}}^{(i)}}\left({\bf{W}}_{{\rm{BS}}}^{(i)}\right)^H{\bf{G}}_{{\rm{B - M}}}^{(i)}} \right)}^{ - 1}} + {{\bf{I}}_{{N_{{\rm{UE}}}}}}} \right]}^{ - 1}}} \right\}
    \end{split}
\end{equation}
\begin{equation}\label{F_Norm}
\begin{split}
\left\| {{{\left( {{\bf{G}}_{{\rm{B - M}}}^{(i)}} \right)}^H}{\bf{W}}_{{\rm{BS}}}^{(i)}} \right\|_{\rm{F}}^2=~&{\rm{tr}}\left\{ {\left( {\tilde{\bf G}}_{{\rm{B-M}}}^{(i)}\right)^H{{\left( {{\bf{B}}_{{\rm{BS}}}^{(i)}} \right)}^{ - 1}}{\bf{Q}}_{{\rm{BS}}}^{(i)}{{\left( {{\bf{B}}_{{\rm{BS}}}^{(i)}} \right)}^{ - 1}}{\tilde{\bf G}}_{{\rm{B-M}}}^{(i)}} \right\}\\
=~&{\rm{tr}}\left\{ {{\bf{{\rm P}}}_{{\rm{BS}}}^{(i)}{\bf{\Sigma }}_{{\rm{BS}}}^{(i)}{{\left( {{\bf{T}}_{{\rm{BS}}}^{(i)}} \right)}^H}{\bf{Q}}_{{\rm{BS}}}^{(i)}{\bf{T}}_{{\rm{BS}}}^{(i)}{{\left( {{\bf{\Sigma }}_{{\rm{BS}}}^{(i)}} \right)}^H}{{\left( {{\bf{{\rm P}}}_{{\rm{BS}}}^{(i)}} \right)}^H}} \right\}
\end{split}
\end{equation}
\end{table*}
Substituting (\ref{Opt2_sepRBS}) into the objective function of (\ref{objective2NEW}), we obtain (\ref{Opt2_Obj2}), which means that minimizing MSE is equivalent to maximizing the term of $\left\| {{{\left( {{\bf{G}}_{{\rm{B - M}}}^{(i)}} \right)}^H}{\bf{W}}_{{\rm{BS}}}^{(i)}} \right\|_{\rm{F}}^2$. Since transmission symbols satisfy ${\rm E}\left\{ {\bf{x}} \right\} = 0$ and ${\rm E}\left\{ {{\bf{x}}{{\bf{x}}^H}} \right\} = {\bf{I}}$, the left-hand in (\ref{constraints22NEW}) equals to ${\rm{tr}}\left\{ {\bf{Q}}_{{\rm{BS}}}^{(i)} \right\}$, where ${\bf{Q}}_{{\rm{BS}}}^{(i)}={{\bf{B}}_{{\rm{BS}}}^{(i)}{\bf{W}}_{{\rm{BS}},2}^{(i)}{{\left( {{\bf{W}}_{{\rm{BS}},2}^{(i)}} \right)}^H}{\bf{B}}_{{\rm{BS}}}^{(i)}}$ with ${\bf{B}}_{{\rm{BS}}}^{(i)}={{\left[ {{\tilde{\bf G}}_{{\rm{B - S}}}^{(i)}{{\left( {{\tilde{\bf G}}_{{\rm{B - S}}}^{(i)}} \right)}^H}} \right]}^{\frac{1}{2}}}$, where ${\tilde {\bf G}}_{{\rm{B - S}}}^{(i)} \buildrel \Delta \over = {\left( {{\bf{W}}_{{\rm{BS}},1}^{(i)}} \right)^H}{\bf{G}}_{{\rm{B - S}}}$ denotes the equivalent channel matrix. Similarly, suppose that ${\tilde{\bf G}}_{{\rm{B - M}}}^{(i)} \buildrel \Delta \over = {\left( {{\bf{W}}_{{\rm{BS}},1}^{(i)}} \right)^H}{\bf{G}}_{{\rm{B - M}}}^{(i)}$, then the objective function becomes (\ref{F_Norm}), where $\left( {\tilde{\bf G}}_{{\rm{B-M}}}^{(i)}\right)^H{\left( {{\bf{B}}_{{\rm{BS}}}^{(i)}} \right)^{ - 1}} = {\bf{{\rm P}}}_{{\rm{BS}}}^{(i)}{\bf{\Sigma }}_{{\rm{BS}}}^{(i)}{\left( {{\bf{T}}_{{\rm{BS}}}^{(i)}} \right)^H}$ with ${\bf{\Sigma }}_{{\rm{BS}}}^{(i)} = {\rm{diag}}\left\{ {\sigma _{{\rm{BS}}}^{(i,1)}, \ldots ,\sigma _{{\rm{BS}}}^{(i,N_{{\rm{UE}}})}} \right\}$ is obtained by SVD in order to further simplify the problem. Using the Hadamard’s inequality (see, e.g., \cite{InforTheory1991}), the optimal solution for maximizing (\ref{F_Norm}) is obtained as ${\bf{Q}}_{{\rm{BS}}}^{(i)*} = {\bf{T}}_{{\rm{BS}}}^{(i)}{\bf{\Lambda }}_{{\rm{BS}}}^{(i)}{\left( {{\bf{T}}_{{\rm{BS}}}^{(i)}} \right)^H}$, where ${\bf{\Lambda }}_{{\rm{BS}}}^{(i)} = {\rm{diag}}\left\{ {\lambda _{{\rm{BS}}}^{(i,1)},\ldots ,\lambda _{{\rm{BS}}}^{(i,N_{{\rm{UE}}})}} \right\}$ with $\lambda _{{\rm{BS}}}^{(i,n)}$~($n=1,\ldots,N_{{\rm{UE}}}$) being the only parameters to be determined. Thus, the objective function in (\ref{Opt2NEW}) can be transformed into $\sum\limits_{n = 1}^{{N_{{\rm{UE}}}}} {{{\left( {\sigma _{{\rm{BS}}}^{(i,n)}} \right)}^2}\lambda _{{\rm{BS}}}^{(i,n)}}$, the constraint in (\ref{constraints21NEW}) is equivalent to $\sum\limits_{n = 1}^{{N_{{\rm{UE}}}}} {\lambda _{{\rm{BS}}}^{(i,n)}}\le \gamma_{\rm {BS}}^{(i)}$, and (\ref{constraints22NEW}) is equivalent to ${\rm{tr}}\left\{ {{\bf{\Lambda }}_{{\rm{BS}}}^{(i)}{\bf{X}}_{{\rm{BS}}}^{(i)}} \right\}=\sum\limits_{n = 1}^{{N_{{\rm{UE}}}}} {x_{{\rm{BS}}}^{(i,n)}\lambda _{{\rm{BS}}}^{(i,n)}} \le \alpha _{{\rm{BS}}}^{(i)}$ with ${\bf{X}}_{{\rm{BS}}}^{(i)} \buildrel \Delta \over =  {\left( {{\bf{T}}_{{\rm{BS}}}^{(i)}} \right)^H}{\left( {{\bf{B}}_{{\rm{BS}}}^{(i)}} \right)^{ - 2}}{\bf{T}}_{{\rm{BS}}}^{(i)}$, where $x_{{\rm{BS}}}^{(i,n)}$ denotes the $(n,n)$-th element of ${\bf{X}}_{{\rm{BS}}}^{(i)}$. In this way, the optimization problem (\ref{Opt2NEW}) can be formulated as
\begin{subequations}
    \label{Opt2NEW1}
    \begin{align}
    \label{objective2NEW1}
    \begin{split}
    \mathop {\min }\limits_{{\bf \lambda} _{{\rm{BS}}}^{(i)}}~& {{{\left( {{\bf c} _{{\rm{BS}}}^{(i)}} \right)}^T}{\bf \lambda} _{{\rm{BS}}}^{(i)}}
    \end{split}\\
    \label{constraints21NEW1}
    \begin{split}
    {\rm{subject~to}}~~&{\bf e}^T{{\bf \lambda} _{{\rm{BS}}}^{(i)}}\le \gamma_{\rm {BS}}^{(i)}
    \end{split}\\
    \label{constraints22NEW1}
    \begin{split}
    &{\left({\bf x}_{{\rm{BS}}}^{(i)}\right)^T{\bf \lambda} _{{\rm{BS}}}^{(i)}} \le \alpha _{{\rm{BS}}}^{(i)}
    \end{split}\\
    \label{constraints23NEW1}
    \begin{split}
    &{-{\bf \lambda} _{{\rm{BS}}}^{(i)}} \le {\bf 0}
    \end{split}
    \end{align}
\end{subequations}
where ${\bf \lambda}_{{\rm{BS}}}^{(i)}=\left[ {\lambda _{{\rm{BS}}}^{(i,1)}, \ldots ,\lambda _{{\rm{BS}}}^{(i,N_{{\rm{UE}}})}} \right]^T$, ${\bf c} _{{\rm{BS}}}^{(i)}=\left[ {-{\left( {\sigma _{{\rm{BS}}}^{(i,n)}} \right)}^2, \ldots ,-\left(\sigma _{{\rm{BS}}}^{(i,N_{{\rm{UE}}})}\right)^2} \right]^T$ and ${\bf x}_{{\rm{BS}}}^{(i)}=\left[ x _{{\rm{BS}}}^{(i,1)},\right.$ $\left.\ldots ,x _{{\rm{BS}}}^{(i,N_{{\rm{UE}}})} \right]^T$. Notably, (\ref{Opt2NEW1}) is a standard linear programming (LP) problem and can easily be solved by CVX. Upon obtaining the optimal ${\bf \lambda}_{{\rm{BS}}}^{(i)*}$, the optimal precoder at the BS is found to be
\begin{equation}
\begin{split}
{\bf{W}}_{\rm{BS}}{\bf{W}}_{{\rm{BS}}}^H=~&{\bf{V}}_{{\rm{BS}}}^{(i,0)}{\left( {{\bf{B}}_{{\rm{BS}}}^{(i)}} \right)^{ - 1}}{\bf{T}}_{{\rm{BS}}}^{(i)}{\bf{\Lambda }}_{{\rm{BS}}}^{(i)*}\\
&\times {\left( {{\bf{T}}_{{\rm{BS}}}^{(i)}} \right)^H}{\left( {{\bf{B}}_{{\rm{BS}}}^{(i)}} \right)^{ - 1}}\left({\bf{V}}_{{\rm{BS}}}^{(i,0)}\right)^H
\end{split}
\end{equation}
from which the optimal ${\bf{W}}_{\rm{BS}}$ can be obtained by SVD.

From the above solution procedure, it is clear that constructing an optimal precoder at the BS only requires the knowledge about the channels from BS to both MUEs and SUEs, a less stringent requirement relative to the sum-MUE minimization based precoding scheme.

\subsection{MSE Minimization at each SC}\label{sec:SCs}
Similarly, the design of precoding vector ${{\bf{W}}_{{\rm{SC}}}^{(s)}}$~($s \in \Omega$) can be handled by solving the LP problem for each SUE, given by
\begin{subequations}
    \label{Opt4NEW1}
    \begin{align}
    \label{objective4NEW1}
    \begin{split}
    \mathop {\min }\limits_{{\bf \lambda} _{{\rm{SC}}}^{(s,j)}}~& {{{\left( {{\bf c} _{{\rm{SC}}}^{(s,j)}} \right)}^T}{\bf \lambda} _{{\rm{SC}}}^{(s,j)}}
    \end{split}\\
    \label{constraints41NEW1}
    \begin{split}
    {\rm{subject~to}}~~&{\bf e}^T{{\bf \lambda} _{{\rm{SC}}}^{(s,j)}}\le \gamma_{\rm {SC}}^{(s,j)}
    \end{split}\\
    \label{constraints42NEW1}
    \begin{split}
    &{\left({\bf x}_{{\rm{SC}}}^{(s,j)}\right)^T{\bf \lambda} _{{\rm{SC}}}^{(s,j)}} \le \alpha _{{\rm{SC}}}^{(s,j)}
    \end{split}\\
    \label{constraints23NEW1}
    \begin{split}
    &{-{\bf \lambda} _{{\rm{SC}}}^{(s,j)}} \le {\bf 0}
    \end{split}
    \end{align}
\end{subequations}
where $j\in J_s$, ${\bf \lambda}_{{\rm{SC}}}^{(s,j)}=\left[ {\lambda _{{\rm{SC}}}^{(s,j,1)}, \ldots ,\lambda _{{\rm{SC}}}^{(s,j,N_{{\rm{UE}}})}} \right]^T$, ${\bf c} _{{\rm{SC}}}^{(s,j)}=\left[ {-{\left( {\sigma _{{\rm{SC}}}^{(s,j,n)}} \right)}^2, \ldots ,-\left(\sigma _{{\rm{SC}}}^{(s,j,N_{{\rm{UE}}})}\right)^2} \right]^T$ and ${\bf x}_{{\rm{SC}}}^{(s,j)}=\left[ x _{{\rm{SC}}}^{(s,j,1)},\ldots, x _{{\rm{SC}}}^{(s,j,N_{{\rm{UE}}})} \right]^T$. Here, the vector elements are calculated accordingly based on the definitions and derivations in Subsection \ref{sec:BS}.

Like the precoder at the BS, only the knowledge about channels from the $s$-th SC to both MUEs and SUEs are required to construct the optimal precoder at the $s$-th SC.

\section{Robust Precoding Design With Imperfect CSI in HetNet}\label{sec:Robust_Prec}
Since perfect CSI is required in the above precoding design, it is often not practical due to channel estimation error, feedback error and quantization error. In this section, we propose more practical precoders for the HetNet with imperfect CSI known at each node.

Assume that the CSI errors of all links are stochastic and modeled as ${\hat{\bf G}}_* = {\bf G}_* + {\bf{\Xi }}_*$, where $*\in\left\{{\rm B-M},{\rm B-S},{\rm S-M},{\rm S-S}\right\}$, ${\hat{\bf G}}_*$ is the estimated channel matrix, and ${\bf{\Xi }}_*$ denotes the channel estimation error matrix which is assumed to be Gaussian distributed with ${\rm{E}}\left\{ {\bf{\Xi }} \right\} = {\bf 0}$ and ${\rm{E}}\left\{ {{\rm vec}\left( {\bf{\Xi }}_* \right){\rm vec}{{\left( {\bf{\Xi }} _*\right)}^H}} \right\} = \sigma _h^2{\bf{I}}$.

\subsection{Robust RAO With Imperfect CSI}\label{sec:Robust_RAO}
With imperfect CSI known at each node, the RAO problem becomes
\begin{subequations}
    \label{Opt1new_imp}
    \begin{align}
    \label{objective1new_imp}
    \begin{split}
    \mathop {\min }\limits_{\begin{array}{*{20}{c}}
    {{{\bf{W}}_{{\rm{BS}}}},{\bf{W}}_{{\rm{SC}}}^{(t)}}\\
    {{{\bf{R}}_{{\rm{BS}}}},{\bf{R}}_{{\rm{SC}}}^{(t)},t \in \Omega }
    \end{array}} &f\left( {{{\bf{W}}_{{\rm{BS}}}},{\bf{W}}_{{\rm{SC}}},{{\bf{R}}_{{\rm{BS}}}},{\bf{R}}_{{\rm{SC}}}} \right)\left| {{\hat{\bf G}}_*} \right.
    \end{split}\\
    \label{constraints1new_imp}
    \begin{split}
    {\rm{subject~to}}~~&(\ref{constraints1new}),~(\ref{constraints2new}),~(\ref{constraints3new})
    \end{split}
    \end{align}
\end{subequations}
where
\begin{equation}\label{Obj_1_imp}
    f\left( {{{\bf{W}}_{{\rm{BS}}}},{\bf{W}}_{{\rm{SC}}},{{\bf{R}}_{{\rm{BS}}}},{\bf{R}}_{{\rm{SC}}}} \right)\left| {{\hat{\bf G}}_*} \right. = {\rm{M\hat S}}{{\rm{E}}_{{\rm{BS}}}} + {\rm{M\hat S}}{{\rm{E}}_{{\rm{SC}}}}
\end{equation}
with (\ref{Obj_ex1_imp}), (\ref{Obj_ex2_imp}),
\begin{table*}
\begin{equation}\label{Obj_ex1_imp}
    \begin{split}
    {\rm{M\hat S}}{{\rm{E}}_{{\rm{BS}}}}  \buildrel \Delta \over = &  {\rm E}\left\{ {{{\left\| {{{\hat{\bf{ x}}}_{{\rm{BS}}}} - {{\bf{x}}_{{\rm{BS}}}}} \right\|}^2}} \left| {{\hat{\bf G}}_*} \right.\right\}={\rm tr}\left\{ {{\bf{R}}_{{\rm{BS}}}}\left[ {{\hat{\bf G}}_{{\rm{B - M}}}^H{{\bf{W}}_{{\rm{BS}}}}{\bf{W}}_{{\rm{BS}}}^H{{\hat{\bf G}}_{{\rm{B - M}}}} + \sum\limits_{s = 1}^S {{{\left( {{\hat{\bf G}}_{{\rm{S - M}}}^{(s)}} \right)}^H}{\bf{W}}_{{\rm{SC}}}^{(s)}{{\left( {{\bf{W}}_{{\rm{SC}}}^{(s)}} \right)}^H}{\hat{\bf G}}_{{\rm{S - M}}}^{(s)}} } \right] \right.\\
    & \left.\times {\bf{R}}_{{\rm{BS}}}^H  - 2{{\bf{R}}_{{\rm{BS}}}}{\hat{\bf G}}_{{\rm{B - M}}}^H{{\bf{W}}_{{\rm{BS}}}} + {{\bf I}_{K{N_{\rm S}}}} + \sigma _0^2{{\bf{R}}_{{\rm{BS}}}}{\bf{R}}_{{\rm{BS}}}^H \right\} + \sigma _h^2\bar \omega{\rm{tr}}\left\{ {{\bf{R}}_{{\rm{BS}}}^H{{\bf{R}}_{{\rm{BS}}}}} \right\}
\end{split}
\end{equation}
\begin{equation}\label{Obj_ex2_imp}
    \begin{split}
    {\rm{M\hat S}}{{\rm{E}}_{{\rm{SC}}}} \buildrel \Delta \over = & {\rm E}\left\{ {\sum\limits_{s = 1}^S {{{\left\| {\hat{\bf{ x}}_{{\rm{SC}}}^{(s)} - {\bf{x}}_{{\rm{SC}}}^{(s)}} \right\|}^2}} } \left| {{\hat{\bf G}}_*} \right.\right\}= \sum\limits_{s = 1}^S {\rm tr}\left\{ {\bf{R}}_{{\rm{SC}}}^{(s)}\left[ {{{\left( {{\hat{\bf G}}_{{\rm{B - S}}}^{(s)}} \right)}^H}{{\bf{W}}_{{\rm{BS}}}}{\bf{W}}_{{\rm{BS}}}^H{\hat{\bf G}}_{{\rm{B - S}}}^{(s)} + \sum\limits_{t = 1}^S {{{\left( {{\hat{\bf G}}_{{\rm{S - S}}}^{(t,s)}} \right)}^H}{\bf{W}}_{{\rm{SC}}}^{(t)}{{\left( {{\bf{W}}_{{\rm{SC}}}^{(t)}} \right)}^H}{\hat{\bf G}}_{{\rm{S - S}}}^{(t,s)}} } \right] \right.\\
    & \left. \times {{\left( {{\bf{R}}_{{\rm{SC}}}^{(s)}} \right)}^H} - 2{\bf{R}}_{{\rm{SC}}}^{(s)}{{\left( {{\hat{\bf G}}_{{\rm{S - S}}}^{(s,s)}} \right)}^H}{\bf{W}}_{{\rm{SC}}}^{(s)} + {{\bf I}_{{L_s}{N_{\rm S}}}} + \sigma _0^2{\bf{R}}_{{\rm{SC}}}^{(s)}{{\left( {{\bf{R}}_{{\rm{SC}}}^{(s)}} \right)}^H} \right\} + \sigma _h^2\bar \omega{\rm{tr}}\left\{ {{{\left( {{\bf{R}}_{{\rm{SC}}}^{(s)}} \right)}^H}{\bf{R}}_{{\rm{SC}}}^{(s)}} \right\}
\end{split}
\end{equation}
\end{table*}
and $\bar \omega = \sum\limits_{t = 1}^S {{\rm{tr}}\left\{ {{{\left( {{\bf{W}}_{{\rm{SC}}}^{(t)}} \right)}^H}{\bf{W}}_{{\rm{SC}}}^{(t)}} \right\}}+ {\rm{tr}}\left\{ {{\bf{W}}_{{\rm{BS}}}^H{{\bf{W}}_{{\rm{BS}}}}} \right\}$. Thus, following the step of RAO, key equations can be derived from the KKT conditions for problem (\ref{Opt1new_imp}) as
\begin{subequations}\label{Derive_23_imp}
    \begin{align}
    \label{item231_imp}
    \begin{split}
    {\bf R}_{\rm{BS}}^{(i)*} = \left( {{\bf{W}}_{{\rm{BS}}}^{(i)}} \right)^H{\hat{\bf G}}_{{\rm{B - M}}}^{(i)}{\left[ \hat{\bf{\Psi }}_{\rm{BS}}^{(i)} +\left( \sigma _0^2+\sigma _h^2\bar\omega\right){\bf{I}}_{N_{\rm{UE}}} \right]^{ - 1}}
    \end{split}\\
    \label{item232_imp}
    \begin{split}
    {\bf R}_{\rm{SC}}^{(s,j)*} = {\left( {{\bf{W}}_{{\rm{SC}}}^{(s,j)}} \right)^H}{\hat{\bf G}}_{{\rm{S - S}}}^{(s,s,j)}{\left[ \hat{\bf\Psi}_{\rm{SC}}^{(s,j)} + \left( \sigma _0^2+\sigma _h^2\bar\omega\right){\bf{I}}_{N_{\rm{UE}}} \right]^{ - 1}}
    \end{split}\\
    \label{WBS_imp}
    \begin{split}
    {\bf{W}}_{{\rm{BS}}}^* = {\left[ {{\hat{\bf{\Phi }}_{{\rm{BS}}}} + \left({\hat\lambda _0}+\sigma_h^2\bar r\right){{\bf{I}}_{{N_{{\rm{BS}}}}}}} \right]^{ - 1}}{\hat{\bf{G}}_{{\rm{B - M}}}}{\bf{R}}_{{\rm{BS}}}^H
    \end{split}\\
    \label{WSC_imp}
    \begin{split}
    {\bf{W}}_{{\rm{SC}}}^{(s)*} = {\left[ {\hat{\bf{\Phi }}_{{\rm{SC}}}^{(s)} + \left({\hat\lambda _s}+\sigma_h^2\bar r\right){{\bf{I}}_{{N_{{\rm{SC}}}}}}} \right]^{ - 1}}\hat{\bf{G}}_{{\rm{S - S}}}^{(s,s)}{\left( {{\bf{R}}_{{\rm{SC}}}^{(s)}} \right)^H}
    \end{split}
    \end{align}
 \end{subequations}
with $i \in I$, $j \in {J_s}$ and $s \in \Omega$, where $\hat{\bf{\Psi}}_{\rm{BS}}^{(i)} = {\left( {\hat{\bf{G}}_{{\rm{B - M}}}^{(i)}} \right)^H}{\bf W}_{\rm{BS}}{\bf{W}}_{\rm{BS}}^H{\hat{\bf G}}_{{\rm{B - M}}}^{(i)} + \sum\limits_{s = 1}^S {{\left( {\hat{\bf G}}_{{\rm{S - M}}}^{(s,i)} \right)}^H}{\bf{W}}_{\rm{SC}}^{(s)}$ ${\left( {{\bf{W}}_{{\rm{SC}}}^{(s)}} \right)}^H{\hat{\bf G}}_{{\rm{S - M}}}^{(s,i)}$, $\hat{\bf{\Psi }}_{{\rm{SC}}}^{(s,j)} = \left( {\hat{\bf G}}_{{\rm{B - S}}}^{(s,j)} \right)^H$ ${\bf{W}}_{{\rm{BS}}}{\bf{W}}_{{\rm{BS}}}^H\hat{\bf G}_{\rm{B - S}}^{(s,j)} + \sum\limits_{t = 1}^S {{{\left( {\hat{\bf G}}_{{\rm{S - S}}}^{(t,s,j)} \right)}^H}{\bf{W}}_{{\rm{SC}}}^{(t)}{{\left( {{\bf{W}}_{{\rm{SC}}}^{(t)}} \right)}^H}{\hat{\bf G}}_{{\rm{S - S}}}^{(t,s,j)}}$, ${\hat{\bf{\Phi }}_{{\rm{BS}}}} = {\hat{\bf{G}}_{{\rm{B - M}}}}{\bf{R}}_{{\rm{BS}}}^H$ ${{\bf{R}}_{{\rm{BS}}}}\hat{\bf{G}}_{{\rm{B - M}}}^H + \sum\limits_{s = 1}^S \hat{\bf{G}}_{{\rm{B - S}}}^{(s)}{\left( {{\bf{R}}_{{\rm{SC}}}^{(s)}} \right)}^H{\bf{R}}_{{\rm{SC}}}^{(s)}{{\left( {\hat{\bf{G}}_{{\rm{B - S}}}^{(s)}} \right)}^H}$, $\hat{\bf{\Phi }}_{{\rm{SC}}}^{(s)} = \hat{\bf{G}}_{{\rm{S - M}}}^{(s)}{\bf{R}}_{{\rm{BS}}}^H{{\bf{R}}_{{\rm{BS}}}}{\left( {\hat{\bf{G}}_{{\rm{S - M}}}^{(s)}} \right)^H} + \sum\limits_{t = 1}^S \hat{\bf{G}}_{{\rm{S - S}}}^{(s,t)}{{\left( {{\bf{R}}_{{\rm{SC}}}^{(t)}} \right)}^H}$ ${\bf{R}}_{{\rm{SC}}}^{(t)}{{\left( {\hat{\bf{G}}_{{\rm{S - S}}}^{(s,t)}} \right)}^H}$, and $\bar r = {\rm{tr}}\left\{ {{\bf{R}}_{{\rm{BS}}}^H{{\bf{R}}_{{\rm{BS}}}}} \right\} + \sum\limits_{t = 1}^S {{\rm{tr}}\left\{ {{{\left( {{\bf{R}}_{{\rm{SC}}}^{(t)}} \right)}^H}{\bf{R}}_{{\rm{SC}}}^{(t)}} \right\}}$.

Based on (\ref{Derive_23_imp}), a robust RAO algorithm can be constructed in a way similar to that in Subsection \ref{constrained} where the RAO algorithm was developed. The optimal Lagrange in the present case satisfy
\begin{subequations}
    \label{LM_imp}
    \begin{align}
    \label{LM_imp1}
    \begin{split}
     \frac{\partial L\left( \hat{\bf{\lambda}} \right)}{\partial {\hat\lambda _0}}&= \sum\limits_{n = 1}^{N_{\rm{BS}}} {\frac{\hat a_{\rm{BS}}^{(n)}}{\left( {\hat d_{\rm {BS}}^{(n)}} + {\hat \lambda _0}+\sigma_h^2\bar r \right)^2}}  - 1 \\
     &\buildrel \Delta \over = {\hat \chi _0}\left( {\hat \lambda _0} \right)= 0
    \end{split}\\
    \label{LM_imp2}
    \begin{split}
     \frac{\partial L\left( \hat{\bf{\lambda}} \right)}{\partial {\hat\lambda _s}} &= \sum\limits_{n = 1}^{N_{\rm{SC}}} {\frac{\hat a_{\rm{SC}}^{(s,n)}}{\left( {\hat d_{\rm {SC}}^{(s,n)}} + {\hat \lambda _s}+\sigma_h^2\bar r \right)^2}}  - 1 \\
     &\buildrel \Delta \over = {\hat\chi _s}\left( {\hat\lambda _s} \right)= 0,~s\in \Omega
    \end{split}
    \end{align}
\end{subequations}
where ${\hat a_{\rm{BS}}^{(n)}}$, ${\hat d_{\rm {BS}}^{(n)}}$, ${\hat a_{\rm{SC}}^{(s,n)}}$ and ${\hat d_{\rm {SC}}^{(s,n)}}$ are defined in an entirely similar way to their counterparts in Subsection \ref{constrained}. Evidently, a bisection search is applicable to (\ref{LM_imp}) to identify the optimal Lagrange multipliers.

\subsection{Robust UAON With Imperfect CSI}\label{sec:Robust_UAON}
As expected, the design of robust UAON with imperfect CSI can be carried out by steps in parallel to those of Algorithm 3. Specifically, the optimal ${\bf R} _{{\rm{BS}}}^*$ and ${\bf R} _{{\rm{SC}}}^{(s)*}$~($s\in \Omega$) have the same expressions as (\ref{item231_imp}) and (\ref{item232_imp}). Consequently, the global minimizers ${{\bf{W}}_{{\rm{BS}}}^*}$ and ${\bf{W}}_{{\rm{SC}}}^{(s)*}~(s\in \Omega)$ for robust UAON can be obtained by first computing
\begin{subequations}
    \label{Derive_25_imp}
    \begin{align}
    \label{W2_imp}
    \begin{split}
    {{{\bf{\bar W}}}_{{\rm{BS}}}} = {\left( {{\hat{\bf{\Phi }}_{{\rm{BS}}}} + \sigma_h^2\bar r{{\bf{I}}_{{N_{{\rm{BS}}}}}}} \right)^{ - 1}}{\hat{\bf{G}}_{{\rm{B - M}}}}{\bf{R}}_{{\rm{BS}}}^H
    \end{split}\\
    \label{ws2_imp}
    \begin{split}
    {\bf{\bar W}}_{{\rm{SC}}}^{(s)} = {\left( {\hat{\bf{\Phi }}_{{\rm{SC}}}^{(s)} + \sigma_h^2\bar r{{\bf{I}}_{{N_{{\rm{SC}}}}}}} \right)^{ - 1}}\hat{\bf{G}}_{{\rm{S - S}}}^{(s,s)}{\left( {{\bf{R}}_{{\rm{SC}}}^{(s)}} \right)^H},~s\in \Omega.
    \end{split}
    \end{align}
\end{subequations}
followed by a norm normalization step as in (\ref{Derive_26}).

\subsection{Robust Non-iterative Algorithm With Imperfect CSI}\label{sec:Robust_Non}
With imperfect CSI, there is a robust counterpart of the non-iterative precoding developed in Subsection~\ref{sec:Sep_Prec} based on separate MSE. To see this, note that the optimal ${\bf{R}}_{{\rm{BS}}}^{(i)}$~($i \in I$) with fixed ${\bf{W}}_{{\rm{BS}}}^{(i)}$ can be expressed as (\ref{Opt2_sepRBS_imp}),
\begin{table*}
\begin{equation}\label{Opt2_sepRBS_imp}
    \begin{split}
    {\bf{R}}_{{\rm{BS}}}^{(i)*} = {\left( {{\bf{W}}_{{\rm{BS}}}^{(i)}} \right)^H}\hat{\bf{G}}_{{\rm{B - M}}}^{(i)}{\left[ {{{\left( {\hat{\bf{G}}_{{\rm{B - M}}}^{(i)}} \right)}^H}{{\bf{W}}_{{\rm{BS}}}^{(i)}}\left({\bf{W}}_{{\rm{BS}}}^{(i)}\right)^H\hat{\bf{G}}_{{\rm{B - M}}}^{(i)} + \left(\sigma _0^2+\sigma_h^2\bar \omega_{\rm BS}^{(i)}\right){{\bf{I}}_{{N_{{\rm{UE}}}}}}} \right]^{ - 1}}
    \end{split}
\end{equation}
\end{table*}
where $\bar \omega_{\rm BS}^{(i)}={\rm tr}\left\{{{\bf{W}}_{{\rm{BS}}}^{(i)}}\left({\bf{W}}_{{\rm{BS}}}^{(i)}\right)^H\right\}$. By substituting (\ref{Opt2_sepRBS_imp}) into the imperfect CSI based objective function, it is evident that minimizing MSE can be transformed into maximizing the term of $\left\| {{{\left( {\hat{\bf{G}}_{{\rm{B - M}}}^{(i)}} \right)}^H}{\bf{W}}_{{\rm{BS}}}^{(i)}} \right\|_{\rm{F}}^2$. In this way, the optimization problem after certain transformations can be rewritten as
\begin{subequations}
    \label{Opt2NEW1_imp}
    \begin{align}
    \label{objective2NEW1imp}
    \begin{split}
    \mathop {\min }\limits_{\hat{\bf \lambda} _{{\rm{BS}}}^{(i)}}~& {{{\left( {\hat{\bf c} _{{\rm{BS}}}^{(i)}} \right)}^T}\hat{\bf \lambda} _{{\rm{BS}}}^{(i)}}
    \end{split}\\
    \label{constraints21NEW1imp}
    \begin{split}
    {\rm{subject~to}}~~&{\bf e}^T{\hat{\bf \lambda} _{{\rm{BS}}}^{(i)}}\le \gamma_{\rm {BS}}^{(i)}
    \end{split}\\
    \label{constraints22NEW1imp}
    \begin{split}
    &{\left(\hat{\bf x}_{{\rm{BS}}}^{(i)}\right)^T\hat{\bf \lambda} _{{\rm{BS}}}^{(i)}} \le \alpha _{{\rm{BS}}}^{(i)}
    \end{split}
    \end{align}
\end{subequations}
where $\hat{\bf \lambda}_{{\rm{BS}}}^{(i)}=\left[ {\hat{\lambda} _{{\rm{BS}}}^{(i,1)}, \ldots ,\hat{\lambda} _{{\rm{BS}}}^{(i,N_{{\rm{UE}}})}} \right]^T$, $\hat{\bf c} _{{\rm{BS}}}^{(i)}=\left[ {-{\left( {\hat\sigma _{{\rm{BS}}}^{(i,n)}} \right)}^2, \ldots ,-\left(\hat\sigma _{{\rm{BS}}}^{(i,N_{{\rm{UE}}})}\right)^2} \right]^T$ and $\hat{\bf x}_{{\rm{BS}}}^{(i)}=\left[ {\hat x} _{{\rm{BS}}}^{(i,1)},\right.$ $\left.\ldots, {\hat x} _{{\rm{BS}}}^{(i,N_{{\rm{UE}}})} \right]^T$. Notably, (\ref{Opt2NEW1_imp}) is a standard linear programming (LP) problem and can easily be solved by CVX. Upon obtaining the optimal $\hat{\bf \lambda}_{{\rm{BS}}}^{(i)*}$, the optimal precoder at the BS is found to be (\ref{Wbs_imp}),
\begin{table*}
\begin{equation}\label{Wbs_imp}
{\bf{W}}_{\rm{BS}}^{(i)}\left({\bf{W}}_{{\rm{BS}}}^{(i)}\right)^H=\hat{\bf{V}}_{{\rm{BS}}}^{(i,0)}{\left( {\hat{\bf{B}}_{{\rm{BS}}}^{(i)}} \right)^{ - 1}}\hat{\bf{T}}_{{\rm{BS}}}^{(i)}\hat{\bf{\Lambda }}_{{\rm{BS}}}^{(i)*}{\left( {\hat{\bf{T}}_{{\rm{BS}}}^{(i)}} \right)^H}{\left( {\hat{\bf{B}}_{{\rm{BS}}}^{(i)}} \right)^{ - 1}}\left(\hat{\bf{V}}_{{\rm{BS}}}^{(i,0)}\right)^H
\end{equation}
\end{table*}
where ${\hat{\bf{B}}_{{\rm{BS}}}^{(i)}}$, $\hat{\bf{T}}_{{\rm{BS}}}^{(i)}$ and $\hat{\bf{V}}_{{\rm{BS}}}^{(i,0)}$ are obtained by replacing all involved ${{\bf G}}_*$ with ${\hat{\bf G}}_*$. Thus, the optimal imperfect CSI based ${\bf{W}}_{\rm{BS}}$ can be obtained by applying SVD to eq.~(\ref{Wbs_imp}). Clearly, constructing an optimal precoder at the BS only requires estimated knowledge about the channels from BS to both MUEs and SUEs, a less stringent requirement relative to the sum-MUE minimization based precoding scheme.

Similarly, the design of precoding vector ${{\bf{W}}_{{\rm{SC}}}^{(s)}}$~($s \in \Omega$) can be handled by solving the LP problem for each SUE, given by
\begin{subequations}
    \label{Opt4NEW1}
    \begin{align}
    \label{objective4NEW1}
    \begin{split}
    \mathop {\min }\limits_{\hat{\bf \lambda} _{{\rm{SC}}}^{(s,j)}}~& {{{\left( {\hat{\bf c} _{{\rm{SC}}}^{(s,j)}} \right)}^T}\hat{\bf \lambda} _{{\rm{SC}}}^{(s,j)}}
    \end{split}\\
    \label{constraints41NEW1}
    \begin{split}
    {\rm{subject~to}}~~&{\bf e}^T{\hat{\bf \lambda} _{{\rm{SC}}}^{(s,j)}}\le \gamma_{\rm {SC}}^{(s,j)}
    \end{split}\\
    \label{constraints42NEW1}
    \begin{split}
    &{\left(\hat{\bf x}_{{\rm{SC}}}^{(s,j)}\right)^T\hat{\bf \lambda} _{{\rm{SC}}}^{(s,j)}} \le \alpha _{{\rm{SC}}}^{(s,j)}
    \end{split}
    \end{align}
\end{subequations}
where $j\in J_s$, $\hat{\bf \lambda}_{{\rm{SC}}}^{(s,j)}=\left[ {\hat\lambda _{{\rm{SC}}}^{(s,j,1)}, \ldots ,\hat\lambda _{{\rm{SC}}}^{(s,j,N_{{\rm{UE}}})}} \right]^T$, $\hat{\bf c} _{{\rm{SC}}}^{(s,j)}=\left[ {-{\left( {\hat \sigma _{{\rm{SC}}}^{(s,j,n)}} \right)}^2, \ldots ,-\left(\hat \sigma _{{\rm{SC}}}^{(s,j,N_{{\rm{UE}}})}\right)^2} \right]^T$ and $\hat{\bf x}_{{\rm{SC}}}^{(s,j)}=\left[ \hat x _{{\rm{SC}}}^{(s,j,1)},\ldots, \hat x _{{\rm{SC}}}^{(s,j,N_{{\rm{UE}}})} \right]^T$. The components of the above vectors are calculated in a way entirely similar to that performed in Sec. IV.A. Here we omit the details due to limited space. Like the precoder at the BS, only estimated knowledge about channels from $s$-th SC to both MUEs and SUEs are required to construct the optimal precoder at $s$-th SC.

\section{Simulation Results}\label{sec:shadowing_simulation}
Simulations were performed for the three MSE-based precoding strategies in the MIMO HetNet systems to demonstrate the efficiency and performance of the proposed precoder design schemes. In the simulations, the bandwidth was 20~MHz, the cell radiuses for macro-cell and small cell were set to 800~m and 100~m, respectively, and the inter site distance between MC and SC was set to 700~m, see Table~\ref{tab:4} for simulation parameters and assumption details. Throughout the simulations, a total of 1000 sets of channel realizations were utilized with each set consisting of $(K+L\times S)$ BS-to-UE channels of size $N_{\rm BS}\times N_{\rm UE}$ and $S\times(K+L\times S)$ SC-to-UE channels of size $N_{\rm SC}\times N_{\rm UE}$, and $10,000$ quadrature-phase-shift keying (QPSK) symbols were transmitted from the BS and each SC node under each channel realization to obtain the BER performance. In all comparisons, unless specified otherwise, the normalized channel estimation error defined by $\bar \sigma_h^2 \buildrel \Delta \over  = {\frac{\sigma_h^2}{\sigma_0^2}}$ was set to be $1$.

\begin{table}[htbp]
\caption{\textsc{Simulation Parameters}}
\label{tab:4}       
\centering
\begin{tabular}[t]{p{90pt}p{140pt}}
\hline\noalign{\smallskip}
Parameters & Setting  \\
\noalign{\smallskip}\hline\noalign{\smallskip}
    Bandwidth & 20~MHz \\
    Cell radius & MC:~800~m, SC:~100~m\\
    Inter site distance & 700~m\\
    Transmit power & BS: $46\sim 56$~dBm, SC: 24~dBm \\
    Noise power density & -174~dBm/Hz  \\
    Number of Antennas & $N_{\rm BS}=36$, $N_{\rm SC}=8$ \\
    Number of UEs & $K=8\sim 18$, $L=4$\\
    Pathloss model (BS) & $\theta_{\rm{BS}}(d)=128.1+37.6\text{log}_{10}(d)$, $d$ (km)~\cite{3GPP2011}\\
    Pathloss model (SC) & $\theta_{\rm{SC}}(d)=140.7+36.7\text{log}_{10}(d)$, $d$ (km)~\cite{3GPP2011}\\
    Penetration loss & $\zeta_{\rm{BS}}=\zeta_{\rm{SC}}=20$~dB~\cite{3GPP2011}\\
  \hline
\end{tabular}
\end{table}

\begin{figure}
   \centering
   \includegraphics[scale=0.53]{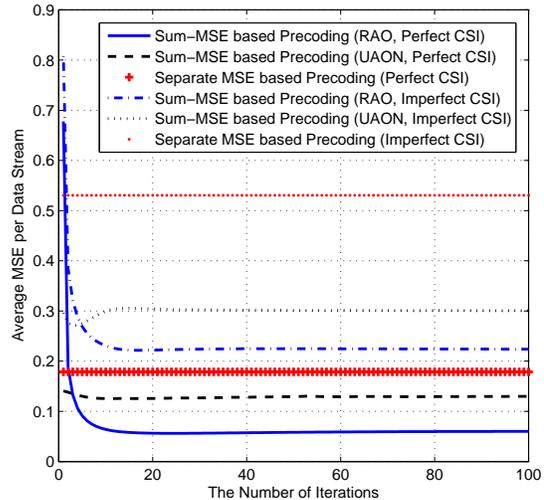}
   \caption{The average MSE per data stream learning curve over 100 runs ($K=8$, ${P_{{\rm{BS}}}}= 46$~dBm, $\bar \sigma_h^2=1$).}
   \label{NumIter_K8}
\end{figure}

Using the proposed sum-MSE based precoding schemes with perfect and imperfect CSI respectively, Fig.~\ref{NumIter_K8} plots the average MSE learning curves over $100$ runs via alternating optimization. For comparison purpose, the red lines in Fig.~\ref{NumIter_K8} depict the average MSE per data stream obtained by the non-iterative algorithm based on separate MSE. From the curves in the figure, it is observed that between the sum-MSE based precoding schemes RAO offers a better performance with a lower average MSE than UAON, but its convergence rate is always slower than the simpler UAON. Also note that the MSE performance of the separate MSE based precoding scheme obtained from the non-iterative algorithm is inferior to that of RAO. Moreover, the performance curves in Fig. 2 reveal that when imperfect CSI is utilized, the MSE differences between the separate MSE based and Sum-MSE based precoding are more pronounced relative to those in the perfect CSI case, for all the three proposed schemes, and the convergence of the robust UAON and robust RAO appears to be slower than their perfect CSI counterpart.

\begin{figure}
   \centering
   \includegraphics[scale=0.495]{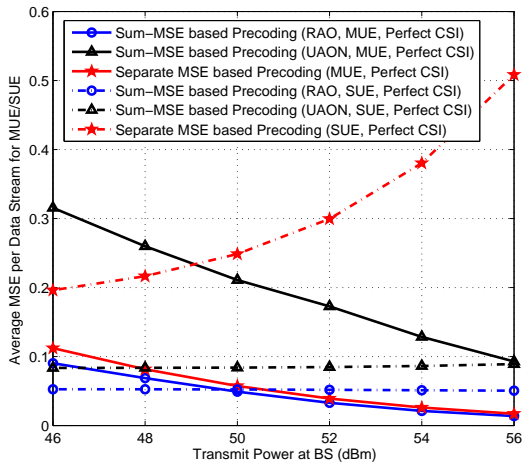}
   \caption{The average MSE per data stream for MUE/SUE versus transmit power at BS ($K=8$, Perfect CSI).}
   \label{MSE_MUE_SUE}
\end{figure}

\begin{figure}
   \centering
   \includegraphics[scale=0.495]{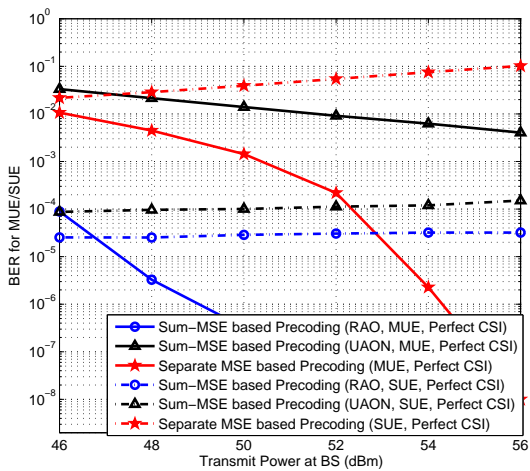}
   \caption{The BER per data stream for MUE/SUE versus transmit power at BS ($K=8$, Perfect CSI).}
   \label{BER_MUE_SUE}
\end{figure}

\begin{figure}
   \centering
   \includegraphics[scale=0.495]{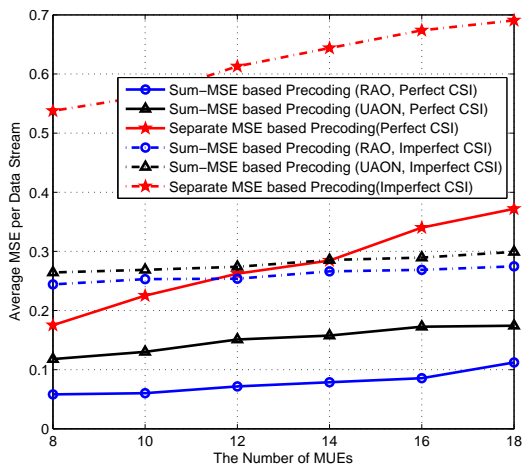}
   \caption{The average MSE per data stream versus the number of MUEs $K$ (${P_{{\rm{BS}}}}= 46$~dBm, $\bar \sigma_h^2=1$).}
   \label{MSE_MUE}
\end{figure}

\begin{figure}
   \centering
   \includegraphics[scale=0.495]{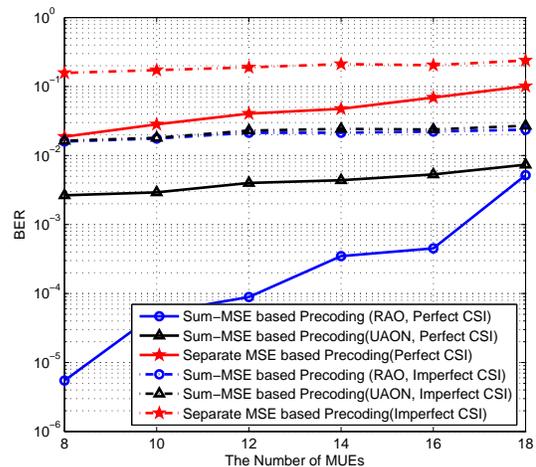}
   \caption{The BER per data stream for MUE/SUE versus the number of MUEs $K$ (${P_{{\rm{BS}}}}= 46$~dBm, $\bar \sigma_h^2=1$).}
   \label{BER_MUE}
\end{figure}

\begin{figure}
   \centering
   \includegraphics[scale=0.495]{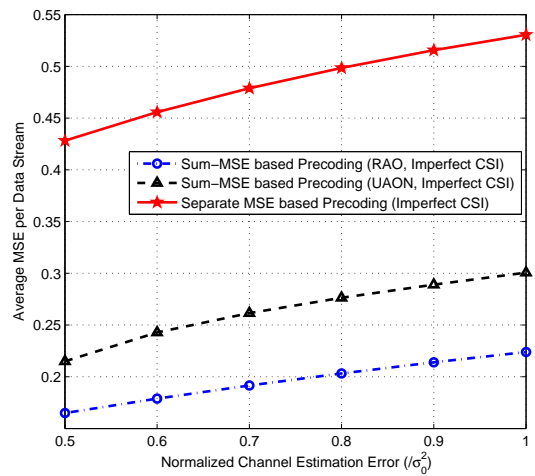}
   \caption{The average MSE per data stream versus normalized channel estimation error $\bar \sigma_h^2$ ($K=8$, ${P_{{\rm{BS}}}}= 46$~dBm, Imperfect CSI).}
   \label{MSE_CE}
\end{figure}

\begin{figure}
   \centering
   \includegraphics[scale=0.495]{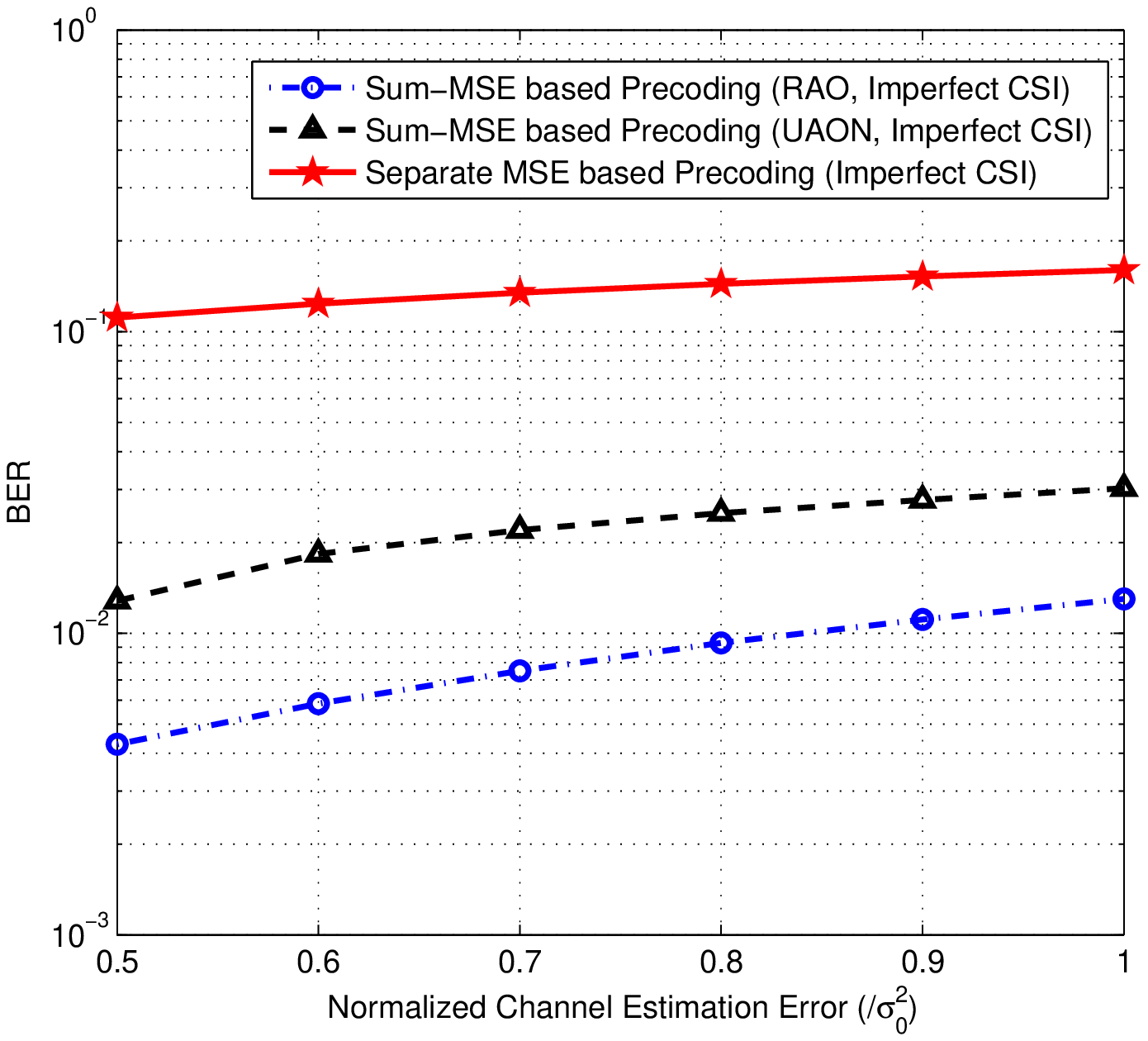}
   \caption{The BER per data stream for MUE/SUE versus normalized channel estimation error $\bar \sigma_h^2$ ($K=8$, ${P_{{\rm{BS}}}}= 46$~dBm, Imperfect CSI).}
   \label{BER_CE}
\end{figure}

Efforts were made to investigate how the average MSE is related to the transmission power. With a fixed ${P_{{\rm{SC}}}}=24$ (dBm), Fig.~\ref{MSE_MUE_SUE} shows that the average MSE for MUEs of iterative algorithms decreases gradually as the transmit power at the BS increases with perfect CSI, while the average MSE for SUEs increases slightly due to the increased inter-cell interferences. Furthermore, the sum-MSE based RAO offers the smallest MSE gap between MUE and SUE, indicating better user fairness, while the separate MSE based precoding has the largest one under the low transmit power. As for the separate MSE based non-iterative precoding scheme, the average MSE for MUEs approaches to that of RAO as the transmit power at BS increases, while the average MSE curve for SUEs goes up gradually. Subsequently, Fig.~\ref{BER_MUE_SUE} illustrates the corresponding BER performance of the proposed schemes, indicating the same relationships as those of the average MSE performance revealed in Fig.~\ref{MSE_MUE_SUE}.

To further illustrate the factors that affect the MSE and BER performance, Fig.~\ref{MSE_MUE} provides the average MSE curves for the three different precoding schemes when the number of MUEs $K$ increases from $8$ to $18$ under both perfect and imperfect CSI cases. It can be seen that the average MSE increases as the number of MUEs gets larger, indicating the higher interferences from other MUEs, and that the sum-MSE based RAO always outperforms both the UAON algorithm and the separate MSE based precoding on the MSE performance under different configurations. Also note that the MSE performance gaps between the separate MSE and sum-MSE based schemes become larger as the number of MUEs increases, i.e., the macro-BS has more antennas relative to the number of the MUEs. Furthermore, the BER performances of the proposed three schemes are given in Fig.~\ref{BER_MUE}, showing a consistent trend with those of Fig.~\ref{MSE_MUE}. We remark that the BER reported here was averaged over all users in the MC and SCs.

In Fig.~\ref{MSE_CE}, the MSE performance of the three proposed schemes with imperfect CSI are depicted versus the normalized channel estimation error $\bar \sigma_h^2$, where $K=8$ and the transmit power at BS was fixed to ${P_{{\rm{BS}}}}= 46$ (dBm). From the figure, it is intuitively clear that the average MSE deteriorates as channel estimation error increases. Similarly, the obtained BER curves Fig.~\ref{BER_CE} are consistent to those in Fig.~\ref{MSE_CE}.


In summary, the sum-MSE based precoding scheme RAO proposed in Section~\ref{sec:Sum_Prec} outperforms the separate MSE based precoding scheme described in Section~\ref{sec:Sep_Prec} in terms of the average MSE per user. On the other hand, RAO requires the information of all channels in the HetNet and its superior performance is achieved at the cost of increased computational complexity relative to that of non-iterative separate MSE based precoding. Furthermore, when the macro-BS has a large number of antennas relative to the number of the MUEs, the performance gap between these two schemes shrinks. As a tradeoff algorithm, the sum-MSE based UAON is much simpler and faster than RAO, with a performance slightly better than the separate MSE based scheme in most configurations with reasonable number of BS antennas.

\section{Conclusion}\label{sec:conclusion}
This paper has developed three new MSE-based precoding schemes for MIMO downlinks in a HetNet architecture consisting of a macro tier overlaid with a second tier of SCs. The first two are both based on the same sum-MSE minimization problem focusing on the joint design of a set of BS and SC transmit precoding matrices or vectors by minimizing the total user MSE under individual transmit power constraints at each cell. On the other hand, we have also proposed a separate MSE minimization based two-level precoder by a non-iterative algorithm in which BD technique is employed as its first-level precoder and each cell designs its own second-level precoder separately without the need to exchange user data or channel state information over the backhaul. On the basis of the estimated imperfect CSI, corresponding robust precoding schemes have been proposed. Simulation results have shown that the sum-MSE based RAO algorithm always outperforms UAON and the separate MSE-based precoding on the MSE performance. When the number of antennas at the macro-BS is large enough relative to the number of MUEs, the average MSE of the low complexity separate MSE-based precoding can come close to those of RAO and UAON. Furthermore, the UAON algorithm has higher convergence rate and lower computation complexity compared to RAO, thus is a worthy trade-off between efficiency and performance.

\section{Proof of Eq. (\ref{Derive_8})}\label{Appendix_B}
To obtain the non-negative multipliers $\lambda_0$ and $\lambda_s$ ($s\in \Omega$) in the above equations, we substitute (\ref{Opt_w}) into (\ref{Derive_4}) and write (\ref{L_Lambda}),
\begin{table*}
\begin{equation}\label{L_Lambda}
    \begin{split}
    &L\left( {\bf{\lambda}} \right) = -{\rm{tr}}\left\{ {{{\left( {{{\bf{\Phi }}_{{\rm{BS}}}} + {\lambda _0}{{\bf{I}}_{{N_{{\rm{BS}}}}}}} \right)}^{ - 1}}{{\bf{G}}_{{\rm{B - M}}}}{\bf{R}}_{{\rm{BS}}}^H{{\bf{R}}_{{\rm{BS}}}}{\bf{G}}_{{\rm{B - M}}}^H} \right\} - {\lambda _0} + \kappa \\
    &~~- \sum\limits_{s = 1}^S {\left[ {{\rm{tr}}\left\{ {{{\left( {{\bf{\Phi }}_{{\rm{SC}}}^{(s)} + {\lambda _t}{{\bf{I}}_{{N_{{\rm{SC}}}}}}} \right)}^{ - 1}}{\bf{G}}_{{\rm{S - S}}}^{(s,s)}{{\left( {{\bf{R}}_{{\rm{SC}}}^{(s)}} \right)}^H}{\bf{R}}_{{\rm{SC}}}^{(s)}{{\left( {{\bf{G}}_{{\rm{S - S}}}^{(s,s)}} \right)}^H}} \right\} + {\lambda _s}}\right]}
    \end{split}
\end{equation}
\end{table*}
where $\kappa  = \sigma _0^2{\rm{tr}}\left\{ {{{\bf{R}}_{{\rm{BS}}}}{\bf{R}}_{{\rm{BS}}}^H} \right\} + \sum\limits_{s = 1}^S {\sigma _0^2{\rm{tr}}\left\{ {{\bf{R}}_{{\rm{SC}}}^{(s)}{{\left( {{\bf{R}}_{{\rm{SC}}}^{(s)}} \right)}^H}} \right\}}  + K{N_{\rm{S}}} + \sum\limits_{s = 1}^S {{L_s}{N_{\rm{S}}}}$ is independent of ${\bf \lambda}$. Then, we start from the expressions of ${{\bf{\Phi}}_{\rm{BS}}}={\bf{S}}_{{\rm{BS}}}^H{{\bf{D}}_{{\rm{BS}}}}{{\bf{S}}_{{\rm{BS}}}}$ where ${\bf{S}}_{{\rm{BS}}}^H{{\bf{S}}_{{\rm{BS}}}} = {{\bf{I}}_{N_{\rm {BS}}}}$ and ${{\bf{D}}_{\rm{BS}}} = {\rm {diag}}\left\{ {{d_{{\rm{BS}}}^{\left( 1 \right)}},{d_{{\rm{BS}}}^{\left( 2 \right)}}, \ldots, {d_{{\rm{BS}}}^{\left( {N_{\rm {BS}}} \right)}}} \right\}$, and ${{\bf{\Phi}}_{{\rm{SC}}}^{(s)}}={\left( {{\bf{S}}_{{\rm{SC}}}^{(s)}} \right)^H}{\bf{D}}_{{\rm{SC}}}^{(s)}{\bf{S}}_{{\rm{SC}}}^{(s)}$~($s\in \Omega$) where ${\left( {{\bf{S}}_{{\rm{SC}}}^{(s)}} \right)^H}{{\bf{S}}_{{\rm{SC}}}^{(s)}} = {\bf{I}}_{N_{\rm {SC}}}$ and ${\bf{D}}_{{\rm{SC}}}^{(s)} = {\rm {diag}}\left\{ {d_{{\rm{SC}}}^{(s,1)}, d_{{\rm{SC}}}^{(s,2)}, \ldots ,d_{{\rm{SC}}}^{(s,{N_{{\rm{SC}}}})}} \right\}$. By substituting the above two expressions into (\ref{L_Lambda}), we obtain (\ref{L_Lambda_simp1}).
\begin{table*}
\begin{equation}\label{L_Lambda_simp1}
    \begin{split}
    &L\left( {\bf{\lambda}} \right)= -{\rm{tr}}\left\{ {{{\left( {{{\bf{D}}_{{\rm{BS}}}} + {\lambda _0}{{\bf{I}}_{{N_{{\rm{BS}}}}}}} \right)}^{ - 1}}{\bf{S}}_{{\rm{BS}}}^H{{\bf{G}}_{{\rm{B - M}}}}{\bf{R}}_{{\rm{BS}}}^H{{\bf{R}}_{{\rm{BS}}}}{\bf{G}}_{{\rm{B - M}}}^H{{\bf{S}}_{{\rm{BS}}}}} \right\}- {\lambda _0} + \kappa \\
    &~~~-\sum\limits_{s = 1}^S {\left[{\rm{tr}}\left\{ {{{\left( {{\bf{D}}_{{\rm{SC}}}^{(s)} + {\lambda _t}{{\bf{I}}_{{N_{{\rm{SC}}}}}}} \right)}^{ - 1}}{{\left( {{\bf{S}}_{{\rm{SC}}}^{(s)}} \right)}^H}{\bf{G}}_{{\rm{S - S}}}^{(s,s)}{{\left( {{\bf{R}}_{{\rm{SC}}}^{(s)}} \right)}^H}{\bf{R}}_{{\rm{SC}}}^{(s)}{{\left( {{\bf{G}}_{{\rm{S - S}}}^{(s,s)}} \right)}^H}{\bf{S}}_{{\rm{SC}}}^{(s)}} \right\} + {\lambda _s}\right]}.
    \end{split}
\end{equation}
\end{table*}
Defining ${\bf A}_{\rm BS} = {{\bf{S}}_{{\rm{BS}}}^H{{\bf{G}}_{{\rm{B - M}}}}{\bf{R}}_{{\rm{BS}}}^H{{\bf{R}}_{{\rm{BS}}}}{\bf{G}}_{{\rm{B - M}}}^H{{\bf{S}}_{{\rm{BS}}}}}$ with the $(n,n)$-th entry denoted as $a_{\rm{BS}}^{(n)}$, and ${\bf A}_{\rm SC}^{(s)} = {\left( {{\bf{S}}_{{\rm{SC}}}^{(s)}} \right)^H}{\bf{G}}_{{\rm{S - S}}}^{(s,s)}{\left( {{\bf{R}}_{{\rm{SC}}}^{(s)}} \right)^H}{\bf{R}}_{{\rm{SC}}}^{(s)}{\left( {{\bf{G}}_{{\rm{S - S}}}^{(s,s)}} \right)^H}$
${\bf{S}}_{{\rm{SC}}}^{(s)}$ with the $(n,n)$-th entry denoted as $a_{\rm{SC}}^{(s,n)}$, we have
\begin{equation}\label{L_Lambda_simp2}
    \begin{split}
    L\left( {\bf{\lambda}} \right)= &-\sum\limits_{n = 1}^{N_{\rm{BS}}} {\frac{a_{\rm{BS}}^{(n)}}{{d_{\rm {BS}}^{(n)}} + {\lambda _0}}} - {\lambda _0}\\
     &- \sum\limits_{s = 1}^S {\left(\sum\limits_{n = 1}^{N_{\rm{SC}}} {\frac{a_{\rm{SC}}^{(s,n)}}{{d_{\rm {SC}}^{(s,n)}} + {\lambda _s} }} + {\lambda _s}\right)} + \kappa.
    \end{split}
\end{equation}
Using (\ref{L_Lambda_simp2}), computing the partial derivative of $L$ w.r.t. $\lambda_0$ and $\lambda_s$~($s\in \Omega$) becomes straightforward, hence the proof of (\ref{Derive_8}).

\end{document}